\documentstyle[12pt]{article}

\begin{document}

{\pagestyle{empty}
\begin{flushright}
SPbU-IP-96-1 \\
hep-ph/96xxxxx
\end{flushright}

\vskip 2cm
\centerline{ \large\bf
         Gauss Integration over Relativistic 3--Body Phase Space}
\centerline{ \large\bf
                for 1--Dimensional Distributions of
                                $  2 \rightarrow 3 $ Reaction }
\vskip 1cm
\centerline{ \bf
              A.A. Bolokhov, P.A. Bolokhov and T.A. Bolokhov }
\centerline{ \it
 St. Petersburg State University, Sankt--Petersburg, Russia}
\centerline{ \bf
                         and }
\centerline{ \bf
                 S.G. Sherman }
\centerline{ \it
St. Petersburg Nuclear Physics Institute,
Sankt--Petersburg, Russia }
\vskip 2cm

\begin{abstract}
                We present the analysis
                of the phase space geometry of
        2 $ \rightarrow $ 3
                reaction
                for the general case of nonzero
                and unequal particle masses.
                Its purpose is to
                elaborate an alternative approach
                to the problem
                of integration over phase space
                which does not exploit
                the Monte Carlo principle.
                The fast and effective algorithm
                of integration
                based on Gauss method is developed
                for treating 1--dimensional distributions
                in two--particle invariant variables.
                The algorithm is characterized by
                significantly improved accuracy
                and it can meet requirements
                of interactive processing.

\end{abstract}

\vskip 6cm
\centerline{
                     Sankt--Petersburg 1996 }

\newpage
}

\vskip 1cm
\centerline{ \bf
         Gauss Integration over Relativistic 3--Body Phase Space}
\centerline{  \bf
                for 1--Dimensional Distributions of
                                $  2 \rightarrow 3 $ Reaction }
\vskip 1cm
\centerline{ \bf
              A.A. Bolokhov, P.A. Bolokhov and T.A. Bolokhov }
\centerline{ \it
 St. Petersburg State University, Sankt--Petersburg, Russia}
\centerline{ \bf
                         and }
\centerline{ \bf
                 S.G. Sherman }
\centerline{ \it
St. Petersburg Nuclear Physics Institute,
Sankt--Petersburg, Russia }
\vskip 1cm

\section
        { Introduction }

                Analysis of kinematics and integration over
                phase space of a reaction are basic problems
                which are common to both experimental and
                theoretical high energy physics.

                In the theoretical field the integration over
                phase space, for example,
                is needed to make theoretical
                predictions confronting the experimental
                measurements and to provide fitting of
                experimental data.
                Besides, a lot of pure theoretical
                investigations such as statistical models of
                nuclear reactions, calculation of unitarity
                corrections, etc., rely upon the integration
                over phase space of final state.

                At all stages of the modern experiments from
                designing the experimental device to
                determination of its characteristics and,
                finally, data treatment, various kinematic
                tools, including integration packages,
                are used for si\-mu\-la\-tion of events,
                determination of acceptances
                and making corrections to the
                measured distributions.

                The Monte--Carlo integration principle
                built
                into the most elaborate GEANT3
\cite{G3}
                system for modeling experimental devices
                provides subroutines for integration purposes
                which are universal in the sense that
                any number of final particles can be treated.
                The time--of--run and precision
                characteristics of the Monte-Carlo based
                programs are not so crucial here since,
                for example, determination of acceptances
                and other device properties is the
                {\it direct} problem.
                At the same time the majority of theoretical
                applications are in the field of
                {\it inverse} problems ---
                to find out physical parameters some kernel
                built of theoretical model and device
                characteristics (angles, momentum cuts, etc.)
                must be inverted when
                applied to the
                experimental results.
                It is well--known that the precision
                requirements for determination of the kernel
                properties which obviously include
                the procedure of integration over
                a domain of phase space
                are much more stringent
                than in the case of the direct problem.

                In contrast with experimental
                applications where
                the integration is usually being performed
                with empty phase space
                the theoretical ones operate with
                complicated model amplitude
                which itself needs considerable
                time for processing.
                The time--of--run properties of the
                integration routines
                become especially vital
                for creating an interactive utilities
                for data bases and routines for data
                treatment
                (like phase shift analysis package SAID
\cite{ArndtFR85}).

                Therefore, development of an alternative
                to Monte--Carlo methods
                of phase space integration even at the
                price of loosing universality in respect
                to the number of particles seems to be
                important.

                Up to now alternative to Monte--Carlo
                methods
                are used only in the simplest case of
    2 $ \rightarrow $ 2
                reactions
                (the phase space is effectively
                1--dimensional and, therefore,
                any integration method, in fact,
                applies well in this occasion).
                Appearing from time to time papers
                like
\cite{BVS89,BardinLR},
                heavily rely on simplification of phase
                space geometry due to equal masses
                or specific amplitude of the reaction.

                Meanwhile the case of the
    2 $ \rightarrow $ 3
                reactions which importance many times
                was stressed since the work
\cite{ChewL59}
                by Chew and Low
                admits almost equally complete elaboration
                as in the
    2 $ \rightarrow $ 2
                case.
                The ground for such a statement
                might be found in the many--years--long
                investigations the results of which are
                summarized in the monographs
\cite{Kopylov70,BK}.
                Of course, the full--scale realization
                of a classical integration scheme
                for all choices of variables of
    2 $ \rightarrow $ 3
                reactions
                and at any kind of kinematical
                constraints will require an approach
                of artificial intelligence and computers
                of extraordinary power.

                The main goal of the present research
                is the demonstration of the principal
                feasibility
                of the classical integration approach.
                This will be done by developing the
                integration algorithm for treating
                total cross sections and 1--dimensional
                distributions in two--particle invariant
                variables
                for the general case of
                unequal particle masses of
    2 $ \rightarrow $ 3
                reaction.
                This problem requires extended analysis
                of the geometry of the phase space of
                the reaction.

                Fortunately,
                almost all the necessary examinations
                and hints
                might be found in various particular chapters
                of the books
\cite{Kopylov70,BK}
                --- one needs to gather carefully crucial
                conclusions and put considerations into
                purposive line of algorithm.
                Certainly, all the key relations
                must be expanded in exact analytic form to
                provide solid programming.
                The presentation of principle answers is an
                incidental goal of the paper.

                Another principle goal is to find a way beyond
                the standard integration algorithms.
                The extremely effective Gauss integration
                method is proved to be quite feasible in the
                considered case.

                The paper is organized as follows.
                Sect. 2 introduces basic notations and the
                starting form of the phase space integral.
                The central Sect. 3 is devoted to
                analysis of necessary
                elements of geometry of the phase space.
                To avoid burst of formulae most of explicit
                expressions are concentrated in Appendix.
                The iterated form of the phase space integral
                is introduced in Sect. 4 where the most
                important properties of the integrated
                expressions are investigated to motivate the
                applicability of Gauss method.
                The specific characteristics of the FORTRAN
                implementation of the results of the previous
                sections are considered in Sect. 5.
                The perspectives of further development of the
                considered approach are discussed in Sect.
                Conclusion.

\newpage

\section { General Definitions }

%

\subsection { Variables }

                We consider the reaction where initial particles
        $ a $
                and
        $ b $
                create 3 particles in the final state.
                We use the notations for momenta of particles
                and for the set of basic variables which are
                very close to that of the book
\cite{BK}.
                They are shown in the diagram

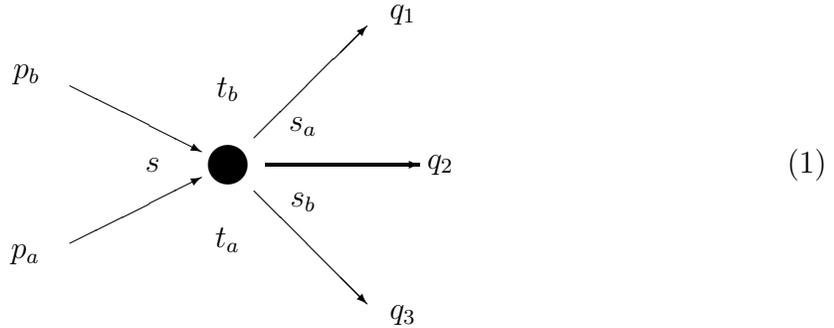
\setcounter{equation}{1}
%
%
\begin{center}
\unitlength=0.50mm
\special{em:linewidth 0.8pt}
\linethickness{0.8pt}
\begin{picture}(320.00,100.00)
\put(160.00,50.00){\circle*{20.00}}
\put(118.00,71.00){\vector(2,-1){35.00}}
\put(118.00,29.00){\vector(2,1){35.00}}
\put(167.00,57.00){\vector(1,1){30.00}}
\put(170.00,50.00){\vector(1,0){41.00}}
\put(167.00,43.00){\vector(1,-1){30.00}}
\put(110.00,74.00){\makebox(0,0)[rc]{$ p_b $}}
\put(110.00,26.00){\makebox(0,0)[rc]{$ p_a $}}
\put(210.00,90.00){\makebox(0,0)[rc]{$ q_1 $}}
\put(220.00,50.00){\makebox(0,0)[rc]{$ q_2 $}}
\put(210.00,10.00){\makebox(0,0)[rc]{$ q_3 $}}
\put(160.00,70.00){\makebox(0,0)[cc]{$ t_b $}}
\put(140.00,50.00){\makebox(0,0)[cc]{$ s $}}
\put(160.00,30.00){\makebox(0,0)[cc]{$ t_a $}}
\put(180.00,60.00){\makebox(0,0)[cc]{$ s_a $}}
\put(180.00,40.00){\makebox(0,0)[cc]{$ s_b $}}
\put(314.00,50.00){\makebox(0,0)[cc]{$ (1) $}}
\end{picture}
\end{center}

                The choice of labels
        $ a $
                and
        $ b $
                is made for convenience of classification
                scheme of the forthcoming integration.
                Momentum transfer
        $ t_a $ ($ t_b $)
                inherits label of the incoming particle.
                For the given particle
        $ a $ ( $ b $ )
                of the diagram
(1)
                there is the pair of associated variables
        $ s, t_a $ ($ s, t_b $)
                (the energy variable
        $ s $
                is assumed to be fixed in the course
                of integration)
                and there exists only one variable
                among the two--particle energies
                of the final state
                which is nonadjacent to the pair
        $ s, t_a $ ($ s, t_b $) --- it gets the same
                label
        $ a $ ( $ b $ ).

                Here,
                we only list the definitions of invariant
                variables
        $ s, s_a, s_b, t_a, t_b $
                in terms of 4--momenta
        $ p_a,p_b,q_1,q_2,q_3 $
                and masses
        $ m_a,m_b,m_1,m_2,m_3 $
                of particles:
\begin{eqnarray}
    s   & = &  (p_a + p_b)^2
               \; = \;  m_a^2 + m_b^2 +
                                2p_a \cdot p_b \; ,
\nonumber \\
    s_a   & = &  (q_1 + q_2)^2
               \; = \;  m_1^2 + m_2^2 +
                                2q_1 \cdot q_2 \; ,
\nonumber \\
    s_b   & = &  (q_2 + q_3)^2
               \; = \;  m_2^2 + m_3^2 +
                                2q_2 \cdot q_3 \; ,
\label{uvar} \\
    t_a   & = &  (p_a - q_3)^2
               \; = \;  m_b^2 + m_3^2 -
                                2p_a \cdot q_3 \; ,
\nonumber \\
    t_b   & = &  (p_b - q_1)^2
               \; = \;  m_b^2 + m_1^2 -
                                2p_b \cdot q_1 \; .
\nonumber
\end{eqnarray}

                The most general case of particle masses
                will be considered:
                all masses are different and nonzero
                --- otherwise considerable simplifications
                are known to take place.

                Every 4-momentum of a particle can enter
                Lorentz-invariant expression only via scalar
                products with another momenta (we consider the
                amplitude of the unpolarized experiment, so
                there are no other 4-vectors for
                con\-struc\-ting
                invariants).

                Five momenta can form 10 scalar products
                provided the mass-shell conditions
\begin{equation}
\label{ms}
  p_a^2 = m_a^2  \; , \;\;
  p_b^2 = m_b^2  \; , \;\;
  q_1^2 = m_1^2  \; , \;\;
  q_2^2 = m_2^2  \; , \;\;
  q_3^2 = m_3^2
\end{equation}
                are fulfilled for all external particles of
                the considered reaction. It is to be noted
                that in a given experiment a specific
                variable, for example, like
        $ ( p_b - p_a )\cdot( q_1 - q_3 ) $,
                might undergo investigation.
                In principle, there is
                combinatorically large amount of
                invariants formed by linear combinations of
                scalar products of momenta of 5 particles.
                Here, we restrict ourselves to the set of
                invariants shown in the diagram
(1).
                We
                note only three features of the quoted
                diagram.

                 First, these very variables, being
                invariant masses of pairs of external
                particles, enter the pole contributions of
                2-particle resonances.
                Therefore, these
                variables are common to almost every
                theoretical analysis.

                Second, the most important property of the
                above variables is their independence. This
                means that any Lorentz-invariant function of
                five 4-momenta of the diagram
(1)
                can be
                written in terms of these 5 invariant
                variables only.
                Avoiding lengthy proof of the independence
                let us
                simply give the explicit expressions of all 10
                products in terms of variables
        $ s,s_a,t_a,s_b,t_b $:
\begin{eqnarray}
             p_b \cdot p_a
             & = &   ( s   - m_b^2 - m_b^2 ) / 2 \; ,
\nonumber \\
             p_b \cdot q_1
             & = & - ( t_b - m_b^2 - m_1^2 ) / 2 \; ,
\nonumber \\
             q_1 \cdot q_2
             & = &   ( s_a - m_1^2 - m_2^2 ) / 2 \; ,
\label{pvar} \\
             q_2 \cdot q_3
             & = &   ( s_b - m_2^2 - m_3^2 ) / 2 \; ,
\nonumber \\
             p_a \cdot q_3
             & = & - ( t_a - m_b^2 - m_3^2 ) / 2 \; ;
\nonumber
\end{eqnarray}
\begin{eqnarray}
\nonumber
      p_b \cdot q_2
             & = & (m_1^2 + m_2^2 - m_3^2 + m_b^2 - m_b^2
\\
\nonumber
             & &
              + 2 q_1 \cdot q_2 - 2 p_b \cdot q_1 + 2 p_a \cdot q_3)/2
                \; ,
\\
\nonumber
      p_b \cdot q_3
             & = & ( - m_1^2 - m_2^2 + m_3^2 + m_b^2 + m_b^2
\\
\nonumber
             & &
               - 2 q_1 \cdot q_2 - 2 p_a \cdot q_3 + 2 p_b \cdot p_a)/2
                \; ,
\\
\nonumber
      p_a \cdot q_1
             & = & (m_1^2 - m_2^2 - m_3^2 + m_b^2 + m_b^2
\\
\label{avar}
             & &
              - 2 q_2 \cdot q_3 - 2 p_b \cdot q_1 + 2 p_b \cdot p_a)/2
                \; ,
\\
\nonumber
      p_a \cdot q_2
             & = & ( - m_1^2 + m_2^2 + m_3^2 - m_b^2 + m_b^2
\\
\nonumber
             & &
               + 2 q_2 \cdot q_3 + 2 p_b \cdot q_1 - 2 p_a \cdot q_3)/2
                \; ,
\\
\nonumber
      q_3 \cdot q_1
             & = & ( - m_1^2 - m_2^2 - m_3^2 + m_b^2 + m_b^2
\\
\nonumber
             & &
               - 2 q_2 \cdot q_3 - 2 q_1 \cdot q_2 + 2 p_b \cdot p_a)/2
                \; .
\end{eqnarray}
                Here, 5 relations of the first group are
                simple inversion of definitions
(\ref{uvar});
                the
                relations of the second group express the rest
                scalar products in terms of ones of the first
                group.

                Third, the
                planar character of the diagram
(1)
                makes it
                convenient to introduce the notion of
                {\it adjacent}
                and
                {\it nonadjacent}
                pairs of variables.
                In the forthcoming analysis one will find a
                principal difference of the geometry of the
                2-dimensional projections of the phase space
                for pairs of adjacent and nonadjacent
                variables.
                In the amplitude analysis the
                difference
                is displayed by the
                fact that variables of nonadjacent pair can
                simultaneously enter the two--pole contribution
                as invariant masses of
                two-particle resonances
                whereas variables of
                adjacent pair never can meet together
                in double-pole  term.

                To conclude the discussion of the choice of
                variables one must notice the following:

                1. Along with the set of variables
(1)
                any
                other set which can be defined in terms of a
                planar diagram will also present the set of
                independent variables.

                2. Performing substitutions of
                particles
        $$ (p_b,p_a,p_1,p_2,p_3) \rightarrow
          (p_{i_a},p_{i_b},p_{i_1},p_{i_2},p_{i_3}) $$
                with
        $ i_a,i_b,i_1,i_2,i_3 $
                being any transposition of the set
        $ \{a,b,1,2,3\} $
                it is possible to use all the results of the
                analysis of a particular case in any other
                planar setting.
                This is provided by invariance
                properties of kinematical functions
                determining the geometry of the phase space.
                These properties are discussed in details in
                the book
\cite{BK}.

                3. However, one can not exploit the full
        $ 5! = 120 $
                transpositions when dealing with
                specific process, for example,
\begin{equation}
        \gamma p \rightarrow \pi^+\pi^0 n \; .
\end{equation}
                Indeed, permutations of particles belonging to
                initial and final states, for example,
        $ \gamma \leftrightarrow \pi^0 $
                or
        $ p \leftrightarrow n $,
                introduces variables for another physical
                processes rather than provides description of
                the considered process in terms of another
                set of variables.

                Therefore, the principal number of different
                sets is
        $ 2! \times 3! = 12 $.

                4. There are many other sets of invariant
                variables which, being strongly nonplanar,
                nevertheless admit the same treatment (an
                example of such a set is:
        $ \{ s, s_a, t_a,
              t_x = ( p_b - p_a ) \cdot ( q_1 - q_2 ),
              t_y = ( p_b + p_a ) \cdot ( q_1 - q_2 )
          \} $).
                Here, our discussion of the principal problem
                of integrating over phase space will be
                restricted to the case of variables described
                by planar diagram
(1).
\subsection{ Phase space integral in invariant variables}

                The end point of theoretical analysis of the
                reaction
(1)
                is the cross section
        $ \sigma (a + b \rightarrow 1 + 2 + 3) $
                which might be written in the form
\begin{equation}
\label{cs}
  \sigma = \sigma_c \frac{f}
                         {4J}
           \int \prod_{j=1}^{3} \frac {d^3 q_j}
                                         {(2\pi)^3 2 q_{j 0}}
           (2\pi)^4 \delta^4 (p_b+p_a-q_1-q_2 -q_3) |M|^2
\end{equation}
                in the case of normalization convention
                adopted, for example, in the book
\cite{DonoghueGH92}.
                Here,
        $ \sigma_c \equiv (\hbar c)^2 =
          0.38937966(23) [{\rm GeV}^2 \, m{\rm barn}] $
                is the conversion constant,
        $ f $ ---
                statistical factor (equal to product of
        $ 1 / n_{\alpha}! $
                over subsets of identical particles) and
\begin{eqnarray}
\label{4J}
   4J & = & 4\sqrt{(p_b \cdot p_a)^2  - m_a^2 m_b^2} =
                2\sqrt{ \lambda(s,m_a^2,m_b^2) }
\\
\label{lam}
     && \left ( \lambda ((p+q)^2,p^2,q^2) \equiv
                 -4 \left |
\begin{array}{cc}
                p^2      & p \cdot q \\
                p\cdot q & q^2
\end{array}
                                \right| \right )
\end{eqnarray}
                stands for normalization of initial state.

                In the papers on particle physics
                different factors in the denominator
                of the integrand
(\ref{cs})
                like
        $ (2\pi)^3 $
                or even
\begin{equation}
\label{2q_0}
         2q_{j 0} \equiv 2\sqrt { m_j^2 + \left |{\bf q}_j\right|^2}
\end{equation}
                sometimes are used to be hidden into the matrix
                element by normalization convention for
                particle states.
                Therefore, the
                definition of the empty phase space
        $ R_3 \equiv R_3 ( |M|^2 = 1 ) $
                is usually based on the common part of
(\ref{cs}):
\begin{equation}
\label{R3}
  R_3 \equiv
           \int \prod_{j=1}^{3} \frac {d^3 q_j}
                                         {2 q_{j 0}}
           \delta^4 (p_b+p_a-\sum_{j}q_j)
\end{equation}
                (energy, or, the same,
        $ s $
                variable is assumed to be fixed in
(\ref{cs})
                and
(\ref{R3})).

                The differential cross sections are always
                experimentally known in terms of
                dis\-tri\-bu\-tions.
                Thus, the integrations in equations
(\ref{cs})
                and
(\ref{R3})
                are assumed to be performed over all region of
                allowed momenta in the case of total cross
                section or over subdomain (slice), cut out
                by bin bounds in the case of
                di\-stri\-bu\-tion.
                The dimension of a di\-stri\-bu\-tion of the
                discussed
        $ 2 \rightarrow 3 $
                process might be
        $ 1,2,3 $
                or
        $ 4 $.

                This directly follows from the counting of
                integration variables in expression
(\ref{R3}):
                there are three particles in the final state;
                their 3-momentum components
        ( $ 9 = 3\times 3 $
                in total ) are integrated while being
                restricted by 4--momentum--conservation
                conditions expressed by
        $ \delta^4 $--
                function ---
        $ 5 = 9-4 $
                degrees of freedom remain.

                Here, we do not consider polarization
                measurements. Therefore, matrix element
        $ | M |^2 $
                has no dependence on the angle in the plane
                orthogonal to the beam axis in the laboratory
                frame where the target particle is at rest.
                Then integration over this variable is easily
                performed providing
        $ 2\pi $
                factor.

                Hence, maximal dimension of a nontrivial
                distribution equals 4.
                The reduction of the 9--dimensional integral
(\ref{cs})
((\ref{R3}))
                to 4-dimensional form in terms of some sets of
                variables of the most interest is discussed in
                details in the book
\cite{BK}.
                In particular, for the case of invariant
                variables
(\ref{uvar})
                the result reads:
        $$
  \sigma =  \frac{\sigma_c f}
                 {2 \sqrt{ \lambda(s,m_a^2,m_b^2) }}
                \frac {1}
                      {(2\pi)^5}
             R_3 ( |M|^2 ); \;
 R_3(|M|^2) = \frac {\pi}
           {4 \sqrt{ \lambda(s,m_a^2,m_b^2) }}
                 r_3(|M|^2) \; ;
        $$
\begin{equation}
\label{R3inv}
 r_3(|M|^2) = \int ds_a dt_a ds_b dt_b \frac {\Theta(-\Delta_4)}
                                         {\sqrt{-\Delta_4}}
                | M |^2   \; .
\end{equation}
                Here,
        $ \Delta_4 $
                is the Gram determinant of any four
                independent momenta, say
        $ p_b,p_a,q_1,q_2 $;
                its explicit form in terms of scalar
                variables
(\ref{uvar})
                is given in the Appendix.
                In what follows we omit subscript 3 in
(\ref{R3inv}), (\ref{cs})
                pointing to the number of particles
                in the final state.

                In practice, the overwhelming majority of
                experimental information on rare processes at
                intermediate energies is represented in the
                form of total cross sections and 1--dimensional
                distributions.
                The reason might be illustrated by
        $ \pi^-p \rightarrow \pi^-\pi^+ n $
                ex\-pe\-ri\-ments
\cite{Blokh63}:
                1023 full--kinematics events constitute solid
                ground for total cross section; 10--14 bins of
                1--dimensional distribution have good filling
                with averaged number of 60--100 events per bin;
                the filling of
        $ 8 \times 8 \ $
                bins of 2--dimensional distribution is
                satisfactory (15 events per bin in average)
                while filling of
        $ 6\times 6\times 6 \ $
                of 3--dimensional ones and
        $ 4\times 4\times 4\times 4 \ $
                4--dimensional bins is poor.

                The increase of statistics of the contemporary
                experiments
\cite{CHAOS624}
                by a factor of 10 considerably improves
                accuracy of results of total cross sections
                and 1--dimensional distributions. When dealing
                with 4--dimensional dist\-ri\-bu\-tions one
                anyway has to make difficult choice between
                poor binning or insufficient filling of
                bins.

                Therefore, the minimal problem which solution
                provides maximal effect is the problem of
                integration over bins of 1--dimensional
                distributions (an overall integration for
                total cross sections is then solved by simple
                summation).

                There is well known property (see, for example,
\cite{BK})
                of the nonadjacent pairs
                (namely,
        $ (s_a, t_a) $
                and
        $ (s_b, t_b)) $
                of variables:
                projections of the phase space onto
                a plane of any such pair admits relatively simple
                description.
                This makes possible to modify
                considerations in such a way that
                integration over 1--dimensional bins might be
                in fact realized by adding up results of
                2--dimensional distribution,
                known as Chew--Low plot:
\begin{eqnarray}
\nonumber
 {
        r_A (|M|^2;j,k)  =
       \rule{5.cm}{0.cm}
 } \\
\label{rChL}
    =  \int ds_a dt_a ds_b dt_b
           \frac {\Theta(-\Delta_4)}
                 {\sqrt{-\Delta_4}}
           | M |^2 \chi (s_a;s_a^{j-1},s_a^{j})
                   \chi (t_a;t_a^{k-1},t_a^{k}) \; ; \\
\nonumber
\\
\nonumber
  {
  \chi(x;x_1,x_2) \equiv  \Theta(x-x_1) \Theta(x_2-x)
                 \; ,
       \rule{2.cm}{0.cm}
  }
\end{eqnarray}
                where
        $ s_a^{j-1}, s_a^{j} $,
        $ t_a^{k-1}, t_a^{k} $
                are bounds of bin
        ($ j , k $)
                in the quoted variables.

                Owing to the symmetry of the phase space in
                the pairs of variables
        $ (s_a, t_a) \leftrightarrow (s_b, t_b) $
                which is induced by
                transpositions of particles:
        $ (a \leftrightarrow b) $,
        $ (1 \leftrightarrow 3) $,
                it is sufficient to perform analysis only
                for the case
(\ref{rChL}).
                To treat 1--dimensional bins in variable of
         $s$--
                (or
         $t$--)
                type one needs to
                rearrange particles in such a way that the
                variable in question becomes
         $s_a$
                (or
         $t_a$)
                variable of the diagram
(1)
                and use trivial binning in the accompanying
                variable of nonadjacent pair ---
         $t_a$
                (or
         $s_a$).

%
%

\newpage

\section  { Geometry of Phase Space }

                The present section deals with the equations
                determining the boundary of the phase space
                over which the integration in eqs.
(\ref{cs}), (\ref{R3inv}), (\ref{rChL})
                has to be performed.
                The main goal is to display the origin of
                expressions for limits of the quoted above
                integrals
                when the latter are written in an appropriate
                successive (iterated) form.
                This requires treating unwieldy formulae
                when they are explicitly expanded.
                Therefore, we collect the most
                cum\-ber\-some
                final expressions in Appendix,
                devoting the discussion of the present section
                to the principal steps of analysis of the
                phase space geometry.

\subsection{ General properties of phase space of two
                non\-ad\-ja\-cent va\-ri\-ab\-les }

                The conditions fixing the 3--particle phase
                space (physical region) of the process
(1)
                might be written in terms of inequalities for
                Gram de\-ter\-mi\-nants
        $ \Delta_n (p_1,...,p_n) $
                built of all possible sets of $n$ momenta of
                particles:
\begin{eqnarray}
        \label{D12}
        \Delta_1 (p_1) \geq 0  \ , \
                \ \Delta_2 (p_1,p_2) \leq 0  \ ; \\
        \label{D3}
        \Delta_3 (p_1,p_2,p_3) \geq 0  \ ;   \\
        \label{D4}
        \Delta_4 (p_1,p_2,p_3,p_4) \leq 0  \ ;  \\
        \label{D5}
        \Delta_5 (p_1,p_2,p_3,p_4,p_5) = 0  \ .
\end{eqnarray}
                Here,
        $\{p_i\}$
                stands for any subset of momenta
        $\{p_a,p_b,q_1,q_2,q_3\}$
                entering diagram
(1).

                The thorough analysis
                of
                how these conditions
                arise and should be treated as well as
                references to original papers might be found
                in
\cite{BK}.
                We only remind that

                $\bullet$
                left hand sides of conditions
(\ref{D12}), (\ref{D3}), (\ref{D4}), (\ref{D5})
                are invariant functions of
                momenta of particles;
                sub\-sti\-tu\-tion of expressions
(\ref{pvar}), (\ref{avar})
                for scalar products of momenta
                shows that eqs.
(\ref{D12}), (\ref{D4}), (\ref{D5})
                are conditions for
        $ s_a, t_a, s_b, t_b $
                at fixed $s$;

                $\bullet$
                conditions
(\ref{D12})
                are in fact statements that all
        $ p_j $
                describe {\it physical} particles with
                nonnegative mass squared
        ($ \Delta_1(p_j) = m_j^2 \geq 0 $)
                and positive energy
        ($ p_{0j} > 0 $);

                $\bullet$
                in the given pattern of iterated
                integration only few of conditions
(\ref{D3})
                are necessary for due treatment (see, for
                example,
\cite{KL});

                $\bullet$
                the choice of momenta in the determinant
        $ \Delta_4 (p_1,p_2,p_3,p_4)$
                has no influence on its value --- it is the
                unique function of independent invariant
                variables;

                $\bullet$
                eq.
(\ref{D5})
                expresses the simple fact that in the
                4--dimensional Min\-kow\-sky space
                there cannot
                be more than 4 linearly independent vectors.


                The condition
(\ref{D4})
                is the natural starting point of analysis.
                The  left hand side of this very condition
                appears in the form
(\ref{R3}), (\ref{rChL})
                of the phase space integral
                when it is rewritten in terms of invariant
                variables.
                The explicit expression of
        $ D_4 \equiv - \Delta_4 ( q_2 , q_3 , p_b , p_a ) $
                is given in the Appendix in terms of expansions
\begin{eqnarray}
\nonumber
 D_4 & = & \sum_{\alpha_s,\alpha_t,\beta_s,\beta_t}
               d_{\alpha_s,\alpha_t,\beta_s,\beta_t}
      s_a^{\alpha_s} t_a^{\alpha_t} s_b^{\beta_t} t_b^{\beta_t} \;
\\
\label{bexp}
     & = & \sum_{\beta_s,\beta_t}
               b_{\beta_s,\beta_t}
                                    s_b^{\beta_t} t_b^{\beta_t} \; .
\end{eqnarray}
                It is reasonable to consider
        $ D_4 $
                as a function of such subset of variables
                in which it takes the simplest form.
                Being generally the form of the fourth degree
                the above
                expression is only  quadratic in any
                pair of nonadjacent variables.
                The total list of such pairs among variables of
                diagram
(1)
                is:
        ($ s_a, t_a $),
        ($ s_b, t_b $)
                and
        ($ t_a, t_b $).
                (Here, we do not list pairs formed with energy
        $ s $
                which is fixed during integration.)

                The pair
        ($ t_a, t_b $)
                can not provide universal treatment since both
                mo\-men\-tum transfers enter the pair
                and only the energy variables
        ($ s_a, s_b $)
                left.
                On the other side the choice of any pair
        ($ s_a, t_a $)
                or
        ($ s_b, t_b $)
                can provide covering of all  cases
                of 1--dimensional distributions.

                Therefore, assuming that
        $ s_a $
                and
        $ t_a $
                acquire some fixed values from the allowed
                domain --- Chew--Low plot
                (the explicit
                description will be given a little later)
                --- it is convenient to write
        $ D_4 $
                in the matrix form
\begin{equation}
\label{D4m}
         D_4 =
                  \left(
\begin{array}{c}
                   s_b \\
                   t_b
\end{array}
                             \right)^{\rm T}
                \cdot \hat{b} \cdot
                 \left(
\begin{array}{c}
                   s_b \\
                   t_b
\end{array}
                             \right)
                + b^{\rm T}  \cdot
                 \left(
\begin{array}{c}
                   s_b \\
                   t_b
\end{array}
                             \right)
                + b_{00} \; ,
\end{equation}
               where
\begin{equation}
                 b \equiv
                 \left(
\begin{array}{c}
                   b_{10} \\
                   b_{01}
\end{array}
                             \right)
                  \; ; \;
                 \hat{b} \equiv
                 \left(
\begin{array}{cc}
                   b_{20} & b_{11}/2 \\
                   b_{11}/2 & b_{02}
\end{array}
                             \right)
                  \; .
\end{equation}

                Let us list the most important properties
                of the form
(\ref{D4m}):

                a) determinant of the matrix
        $ \hat{b} $
\begin{eqnarray}
         {\rm Det} \; \hat{b}  =
          s_a
            D_{3a}
        / 16  \; ,
\label{Dtbf}
\end{eqnarray}
                where explicit expression for determinant
       $
          D_{3a}  \equiv  \Delta_3 ( q_3 , p_b , p_a )
            \;
       $
                is given by eq.
(\ref{D3f})
                of Appendix,
                is nonnegative in the physical region
                since
                two--particle energy
        $ s_a $
                is positive and for the Gram determinant
        $ D_{3a} $
                condition
(\ref{D3})
                is valid;

                b) trace of the matrix
        $ \hat{b} $
                is nonpositive  --- analyzing the
                diagonal elements
        $ b_{02} $, $ b_{20} $
                of
        $ \hat{b} $
                provided by eqs.
(\ref{bxx})
                one can recognize
        $ \lambda $--function
                expressions for combinations of momenta
                for which the conditions
(\ref{D12})
                are fulfilled;

                c) when
        $ D_4 $
                is transformed to the centered form
(\ref{D4c})
                the free term
        $ b_c $
                of the latter
\begin{equation}
         b_c \equiv  b_{00}
                  \;  -  \;
                  {{1}\over{4}}
                 b^{\rm T} \cdot \hat{b}^{-1} \cdot b
                  \;  =  \;
           D_{2a}  D_{3a}
             /  s_a  \; .
\label{bcf}
\end{equation}
                is nonnegative
                in the physical domain of
        $ s_a $, $ t_a $
                variables
                --- this is direct consequence of conditions
(\ref{D12}), (\ref{D3})
                applied to the RHS of eq.
(\ref{bcf});

                d) whenever determinant
        $ | \hat{b} | $
                becomes zero
        $ b_c $
                vanishes also;
                the ratio
        $ b_c / {\rm Det} \; \hat{b} $
                remains finite in the physical domain
                --- by comparing
(\ref{bcf})
                and
(\ref{Dtb})
                one has
\begin{equation}
\label{bcdtbrat}
                 b_c / ( {\rm Det} \; \hat{b} ) =
                        16 D_{2a} / s_a^2
                \; .
\end{equation}

                These properties imply that condition
(\ref{D4})
                determines an ellipse in the
        ($ t_b , s_b $)
                plane.
                Before proceeding with its analysis let
                us recall that
        $ t_a $ and $ s_a $
                variables are assumed to be fixed.
                The location of the
        ($ t_b , s_b $)
                ellipse and orientation
                of its principal axes strongly depends on
                the latter variables:
                this is demonstrated by the Fig. 1
                where families of ellipses are drawn in the
        ($ t_b , s_b $)
                plane
                for different values of
        $ t_a $, $ s_a $
                from the allowed domain.
                Horizontal sequences of elliptic curves
                are obtained for a fixed value of
        $ s_a $
                --- from eq.
(\ref{sbc})
                given in the Appendix
                one can see that
        $ s_b^c $
                coordinate of the center of the ellipse
                does not depend on
        $ t_a $
                variable.

\subsection{ Boundaries of phase space for
                non\-ad\-ja\-cent va\-ri\-ab\-les }

                The standard way to derive the
                bounds for the
        ($ t_b , s_b $)
                variables is to solve first the equation
        $ D_4 = 0 $
                for a single--variable form
(\ref{D4bs})
((\ref{D4bt}))
                which is scanned in the Appendix:
\begin{equation}
\label{sblr}
          s_b^{l,r} ( t_b )
                  \;  =  \;
               \frac{ (- b_{s1} \pm \sqrt { b_s } ) }
                           { 2 b_{s2} }
            \; ;
\end{equation}
\begin{equation}
\label{tblr}
          t_b^{l,r} ( s_b )
                  \;  =  \;
               \frac{ (- b_{t1} \pm \sqrt { b_t } ) }
                           { 2 b_{t2} }
            \; .
\end{equation}
                Here and in what follows we use the
                pattern of subscript mnemonics
                ({\sl L}eft--{\sl R}ight),
                ({\sl D}own--{\sl U}p),
                ({\sl G}round--{\sl H}igh)
                to mark the solution.
                Small letters will be used for
                boundary functions
                (like
        $ (l, r) $
                in eqs.
(\ref{sblr}), (\ref{tblr}))
                whereas capital ones ---
                for absolute bounds of variables in the
                considered plot.

                The above solutions
                by virtue of relations
(\ref{bs20}), (\ref{bt20})
                (see subsect. 2.2 of Appendix)
                are expressed
                in terms of coefficients
        $ b_{\beta_s,\beta_t} $
                and discriminants
        $ b_s, b_t $.
                The latter
                are also written out explicitly
                in eqs.
(\ref{dscbs}), (\ref{dscbt})
                and
(\ref{D3Bs}), (\ref{D3Bt})
                of Appendix.
                Note that the plus sign at square root in
(\ref{tblr}) ((\ref{sblr}))
                corresponds to the {\it left} bound of
        $ t_b $ ($ s_b $)
                because the denominator
        $ b_{t2} = b_{02} $ ($ b_{s2} = b_{20} $)
                is already stated to be negative.

                Being the product of factors
        $ D_{3a} $ and $ D_{3as} $ ($ D_{3a} $ and $ D_{3at} $)
                discriminant
        $ b_s $  ($ b_t $)
                vanish, first, when
        $ D_{3a} = 0 $
                --- this condition determines the
                absolute bounds of
        $ t_a $, $ s_a $
                pair of variables;
                it will be discussed a little later ---
                and, second, when
        $ D_{3as} = 0 $ ($ D_{3at} = 0 $).
                The solutions of the latter provide
                bounds
        $ t_b^{L} $, $ t_b^{R} $
        ($ s_b^{L} $, $ s_b^{R} $)
                of interval spanned by
        $ t_b $
        ($ s_b $)
                variable at given values of
        $ t_a $ and $ s_a $:
\begin{equation}
\label{tbLR}
          t_b^{L,R}
                  \;  =  \;
               \frac{ (- b_{st1} \pm \sqrt { b_{st} } ) }
                           { 2 b_{st2} }
            \; ;
\end{equation}
\begin{equation}
\label{sbLR}
          s_b^{L,R}
                  \;  =  \;
               \frac{ (- b_{ts1} \pm \sqrt { b_{ts} } ) }
                           { 2 b_{ts2} }
            \; .
\end{equation}
                The explicit expressions for expansion
                coefficients
         $ b_{st\beta} $  and $ b_{ts\beta} $
                of
        $ D_{3as} $ and $ D_{3at} $
                as well as for discriminants
        $ b_{ts} $  and $ b_{st} $
                of these forms are given in Appendix.
                The important issue is the presence of
                the factor
        $ D_{2a} $
                in both discriminants
                (cf. eqs.
(\ref{dstf}), (\ref{dtsf})
                and
(\ref{D2aa})).

                Analysis of the above discriminants and the
                properties a)--c) of the form
(\ref{D4m})
                of the previous subsection shows
                that all the
                {\it necessary} and {\it sufficient}
                conditions
                for existing a nondegenerate domain in
        ($ t_b , s_b $)
                plot
                are collected in the requirement that
                free term
        ($ b_c $)
                is positive.
                Because of restrictions
(\ref{D12}), (\ref{D3})
                the only unambiguous splitting of the
                product
(\ref{bcf})
                is
        $$
           D_{2a} \; > \; 0 \; ; \; \; D_{3a}  \; > \; 0 \; .
        $$

                These conditions determine
                the Chew--Low plot in
        $ s_a , t_a $
                variables.
                Its boundaries correspond to curves in
        ($ s_a , t_a $)
                plane
                where LHS's of the above inequalities
                vanish.
                Because of factorization of the expression
(\ref{D2aa})
                for Gram determinant
        $ D_{2a} $
                it is easy to see that one of the boundary
                curve is simply the straight line
\begin{equation}
\label{saLline}
            s_a - ( m_1 + m_2)^2
                  \;  =  \;
                      0
            \;
\end{equation}
                and the other line provided by
        $ D_{2a} $
\begin{equation}
\label{saLf}
            s_a - ( m_1 - m_2)^2
                  \;  =  \;
                      0
            \;
\end{equation}
                is well out of phase space
                for the square of two--particle energy.

                Here it is where an asymmetry between the
                momentum transfer
        $ t_a $
                and energy variable
        $ s_a $
                appears.
                The lower (left) absolute bound
        $ s_a^{0} $
                for the latter does not depend on
                initial energy
        $ s $:
\begin{equation}
\label{sa0f}
            s_a^{0}
                  \;  =  \;
                        ( m_1 + m_2)^2
            \;  .
\end{equation}

                Another boundary curve in
        ($ s_a , t_a $)
                plane is provided by vanishing of
                the
        $ D_{3a} $
                factor of
        $ b_c $.
		This factor, being the second order form
                of the
        $ s_a , t_a $
		variables,
                is presented in Appendix in the
                manner
                similar to that
                of
        $ D_4 $
                case.
                Now, the quadratic form
\begin{equation}
\label{D3Af}
         D_{3a} =
                  \left(
\begin{array}{c}
                   s_a \\
                   t_a
\end{array}
                             \right)^{\rm T}
                \cdot \hat{A} \cdot
                 \left(
\begin{array}{c}
                   s_a \\
                   t_a
\end{array}
                             \right)
                + A^{\rm T}  \cdot
                 \left(
\begin{array}{c}
                   s_a \\
                   t_a
\end{array}
                             \right)
                + A_{00} \; ,
\end{equation}
               where
\begin{equation}
                 A \equiv
                 \left(
\begin{array}{c}
                   A_{10} \\
                   A_{01}
\end{array}
                             \right)
                  \; ; \;
                 \hat{A} \equiv
                 \left(
\begin{array}{cc}
                   A_{20} & A_{11}/2 \\
                   A_{11}/2 & A_{02}
\end{array}
                             \right)
                  \;
\end{equation}
                are determined by coefficients provided
                by eqs.
(\ref{axx})
                of Appendix,
                is of hyperbolic type.
		The conclusion is not so difficult
                to arrive to,
                basing on the properties of this form
                presented in Appendix.
                In particular,
                the roots in a variable
                (provided the accompanying variable
                is fixed) are
\begin{equation}
\label{safh}
          s_a^{g,h} ( t_a )
                  \;  =  \;
               \frac{ ( - ( A_{11} t_a + A_{10} )
                                \pm \sqrt { A_s } ) }
                           { 2 A_{20} }
            \; ;
\end{equation}
\begin{equation}
\label{tagh}
          t_a^{g,h} ( s_a )
                  \;  =  \;
               \frac{ ( - ( A_{11} s_a + A_{01} )
                                \pm \sqrt { A_t } ) }
                           { 2 A_{02} }
            \; ,
\end{equation}
                where discriminants
        $ A_s $, $ A_t $
                are
\begin{eqnarray}
\label{As}
                A_s & \equiv & ( A_{11} t_a + A_{10} )^2
                          - 4 A_{20}
                ( A_{02} t_a^2 + A_{01} t_a + A_{00} )
\\
 & = &
\nonumber
                          D_{2c} D_{2at} =
                          D_{2c} \,
  (t_a - ( m_a + m_3 )^2 )
  (t_a - ( m_a - m_3 )^2 )
	       / 4
\; ;
\end{eqnarray}

\begin{eqnarray}
\label{At}
                A_t & \equiv & ( A_{11} s_a + A_{01} )^2
                          - 4 A_{02}
                ( A_{20} s_a^2 + A_{10} s_a + A_{00} )
\\
 & = &
\nonumber
                          D_{2c} D_{2as} =
                          D_{2c} \,
  (s_a - ( \sqrt{s} + m_3 )^2 )
  (s_a - ( \sqrt{s} - m_3 )^2 )
	       / 4
\; .
\end{eqnarray}

                This helps to fix critical points
        $ t_a^{G,H} $, $ s_a^{G,H} $
                of the considered variables as roots of
                the above discriminants:
\begin{equation}
\label{taLR}
          t_a^{G,H}
                  \;  =  \;
                        ( m_a \mp m_3 )^2
            \; ;
\end{equation}
\begin{equation}
\label{saLR}
          s_a^{G,H}
                  \;  =  \;
                        ( \sqrt{s} \mp m_3 )^2
            \; .
\end{equation}

                Unlike the previous case of second--order
                form
        $ D_4 $
                in
        $ s_b $, $ t_b $
                variables,
                here, the intervals between roots
                are nonphysical since
        $ D_{3a} < 0 $
                there.
                Regions
        $ t_a > t_a^H  $
                and
        $ s_a > s_a^H  $
                are cross regions of another processes:
        $ a + 3 \rightarrow b + 1 + 2  $
                and
        $ a + b + 3 \rightarrow 1 + 2  $
                respectively.
                It is easy to see that
                the absolute range of variation of
        $ s_a $
                variable is fixed in an unique way by eqs.
(\ref{sa0f})
                and
(\ref{saLR}):
\begin{equation}
\label{sarange}
      ( s_a^{D} \equiv  s_a^0 )
                <  s_a   <
                        ( s_a^{U} \equiv  s_a^{G} )
            \; .
\end{equation}
                For a given
        $ s_a $
                bounds
        $ t_a^{d,u} ( s_a )  $
                of
        $ t_a $
                interval are those given by eq.
(\ref{tagh})
\begin{equation}
\label{tadu}
         t_a^{d,u} ( s_a )  =
         t_a^{g,h} ( s_a )
            \; .
\end{equation}

                The absolute bounds of momentum transfer
        $ t_a $
                are provided in a different way.
                It is simple to find that the lowest
                possible value of
        $ t_a $
                is given by the lowest intersection point
                of hyperbola
        $ D_{3a} = 0 $
                with the line
        $ s_a = s_a^0 $:
\begin{equation}
\label{taD}
          t_a^D   =  t_a^g ( s_a^0 )
            \; .
\end{equation}
                For determination of
                the upper bound it is crucial whether
                the common point
\begin{equation}
         s_a^T \equiv \{ s (m_a - m_3)
                + m_3 (m_3 m_a + m_b^2 - m_a^2 )
                 \}       / m_a
\end{equation}
                of the tangent line
        $ t_a = t_a^D $
                and the hyperbola
                belongs to the physical region
(\ref{sarange})
                of
        $ s_a $
                variable or is located below.
                In the former case the value
        $ t_a^G $
                is never attained.
                The discussed condition depends on
                the particular relations of the
                particle masses (and energy region);
                the relevant classification might
                be found in the paper
\cite{KL}.
                For the purpose of calculations
                it is sufficient simply to compare
        $ s_a^T $
                with
        $ s_a^0 $.
                Then the upper bound
        $ t_a^U $
                is given by eqs.
\begin{equation}
\label{taUif}
                {\sf if }
                        \; \;
                        s_a^T < s_a^0
                        \; \;
                {\sf then }
                        \; \;
                        t_a^U = t_a^h ( s_a^0 )
\end{equation}
\begin{equation}
\label{taUelse}
                {\sf else }
                        \; \;
          t_a^U = t_a^G
           \; .
\end{equation}

                The behavior of the
        $ s_a $
                bounds
        $ s_a^{d,u} $
                at given
        $ t_a $
                follows the similar scheme:
\begin{equation}
\label{sau}
                        s_a^u ( t_a ) = s_a^h ( t_a ) \; ; \;
\end{equation}
\begin{equation}
\label{sadif}
                {\sf if }
                        \; \;
                        t_a < t_a^h ( s_a^0 )
                        \; \;
                {\sf then }
                        \; \;
                        s_a^d ( t_a ) =  s_a^0  \;
\end{equation}
\begin{equation}
\label{sadelse}
                {\sf else }
                        \; \;
                        s_a^d ( t_a ) = s_a^g ( t_a )
           \; .
\end{equation}

                The variety of labels is dictated by the
                need of avoiding overlappings
                when re\-pro\-du\-cing
                for the case of
        ($ s_b, t_b $)--plot.
                The symmetry
\begin{equation}
\label{absymm}
        a \leftrightarrow b \; , \; 1 \leftrightarrow 3
\end{equation}
                of diagram
(1)
                allows to use all the formulae of the
                current subsect. with the substitution
(\ref{absymm})
                made for all indices and labels
                containing
        $ \{ a, b, 1, 2, 3 \} $.

                To resume this Sect. one should
                note that the symmetry in two--particle
                energy and momentum transfer variables
                which is perfect for
        $ s_b, t_b $
                at fixed
        $ s_a, t_a $
                (eqs.
(\ref{sblr}), (\ref{tblr}),
(\ref{sbLR}), (\ref{tbLR})
                following from the symmetric representation
(\ref{D4m})
                of boundary function
        $ D_4 $)
                is broken in the resulting description of
                Chew--Low plot:
(\ref{sarange}), (\ref{tadu})
                versus
(\ref{taUif}--\ref{sadelse}).
                This must be carefully implemented into
                algorithms performing integration over
                phase space.

\newpage

\section { Gauss Integration }

\subsection{ Iterated form of phase space integral }

                The analysis of the phase space geometry
                provided by the previous section
                makes it possible to write
                the phase space integral
(\ref{R3inv})
((\ref{rChL}))
                in a number of iterated forms.
                There is
                the natural splitting of 4--dimensional
                integral into internal integral over
        $ s_b , t_b $
                variables
\begin{equation}
\label{rb}
   r_b (|M|^2)
    =  \int ds_b dt_b
           \frac {\Theta(-\Delta_4)}
                 {\sqrt{-\Delta_4}}
           | M |^2
                \;
\end{equation}
                and the integral over the rest ones,
                the final form of
(\ref{rChL})
                being
\begin{equation}
\label{rA}
 {  r_A (|M|^2; j, k )  = }
      \int ds_a dt_a
           \frac {\Theta(-\Delta_4)}
                 {\sqrt{-\Delta_4}}
                   \chi (s_a;s_a^{j-1},s_a^{j})
                   \chi (t_a;t_a^{k-1},t_a^{k})
                     r_b (|M|^2)
                \; .
\end{equation}

                Basing on the calculated boundary functions
(\ref{sblr}),
(\ref{tblr}),
(\ref{sbLR}),
(\ref{tbLR})
                the internal integral might be written in
                two iterated forms, namely:
\begin{equation}
\label{rbst}
   r_b (|M|^2) =
      \int_{s_b^L}^{s_b^R} ds_b
      \int_{t_b^l (s_b)}^{t_b^r (s_b)} dt_b
           \frac {\Theta(-\Delta_4)}
                 {\sqrt{-\Delta_4}}
           | M |^2
                \;
\end{equation}
                and
\begin{equation}
\label{rbts}
   r_b (|M|^2) =
      \int_{t_b^L}^{t_b^R} dt_b
      \int_{s_b^l (t_b)}^{s_b^r (t_b)} ds_b
           \frac {\Theta(-\Delta_4)}
                 {\sqrt{-\Delta_4}}
           | M |^2
                \; .
\end{equation}
                There are no reasons to prefer
                one form or another basing on pure
                geometric arguments --- the
        $ s_b , t_b $
                domain was stated to be quite symmetric.
                However, the integrated amplitude
        $ M $
                might have different behavior in the
                discussed variables.
                For example, the amplitude might have
                poles in two--particle energy
        $ s_b $
                (shifted to complex plane from the real axe)
                and cuts
                whereas there should be no such reason
                of rapid variation with momentum transfer
        $ t_b $
                in the integration domain.
                Since there is maximal step limit
                among the parameters terminating the
                calculations of integrating procedures
                it is reasonable to perform the integration
                over "smooth" variable
        $ t_b $
                first --- otherwise all the rest integrations
                are expected to be performed at maximal step
                number as well.

                As to external integral
(\ref{rA})
                the difference in the description of
        ($ s_a , t_a $)--plot
                in the variables
                which was discussed at the
                end of previous section
                makes it more convenient to choose the order
\begin{eqnarray}
\label{rAst}
 {
       r_A (|M|^2; j, k)  =
       \rule{4.cm}{0.cm}
 }
\\
\nonumber
      \int_{s_a^D}^{s_a^U} ds_a
      \int_{t_a^d (s_a)}^{t_a^u (s_a)} dt_a
           \frac {\Theta(-\Delta_4)}
                 {\sqrt{-\Delta_4}}
                   \chi (s_a;s_a^{j-1},s_a^{j})
                   \chi (t_a;t_a^{k-1},t_a^{k})
                     r_b (|M|^2)
                \;
\end{eqnarray}
                to simplify the program logic.
                The order
\begin{eqnarray}
\label{rAts}
 {
        r_A (|M|^2; j, k)  =
       \rule{4.cm}{0.cm}
 }
\\
\nonumber
      \int_{t_a^D}^{t_a^U} dt_a
      \int_{s_a^d (t_a)}^{s_a^u (t_a)} ds_a
           \frac {\Theta(-\Delta_4)}
                 {\sqrt{-\Delta_4}}
                   \chi (s_a;s_a^{j-1},s_a^{j})
                   \chi (t_a;t_a^{k-1},t_a^{k})
                     r_b (|M|^2)
                \;
\end{eqnarray}
                which directly provide
                1--dimensional distribution
                in momentum transfer
        $ t_a $
                is less attractive from the point of
                view of amplitude behavior and,
                what is more important,
                it causes considerable complications
                of the program logic
                which must trace all details of
                conditions
(\ref{taUif}--\ref{sadelse}).
                Because of these complications
                the program for calculation of
                1--dimensional
        $ t_a $
                distribution based on representation
(\ref{rAts})
                is found to be equivalent to the
                program for processing
                2--dimensional distribution
                according to eq.
(\ref{rAst}).

                When we have written the explicit form
(\ref{rAst}),
(\ref{rbst})
                of the phase space integral
(\ref{rChL}),
                the problem of calculation of all
                1--dimensional distributions in
                two--particle invariant variables
                is, in principle, solved by an
                universal procedure
                which is designed in a
                straightforward way
                according to the discussed formulae
                (and boundary eqs. of the previous
                section
                written in terms of quantities
                explicitly given in the Appendix).
                Even with the use of integration
                routines based on Simpson or other
                standard algorithm
                the procedure is much more effective
                when compared to the Monte Carlo based
                analog.

                In the rest part of the current section
                we shall consider the implementation
                of the Gauss integration method for
                the integral in question.
                The possibility of further improvement
                of characteristics of the program
                comes from the following observations:

                1) the presence of denominator in eq.
(\ref{rb})
                which vanish at the boundary of the
                phase space
                makes it evident that the standard
                integrating routines
                waste
                almost all
                the time processing this (integrable)
                singularity
                (one can avoid the difficulty by a change of
                variables.
                This results in minor complications
                of transfer of arguments from the integrating
                procedure to the user amplitude.
                It is the loss of universality for further
                development of algorithms
                processing 3-- and 4-- dimensional distributions
                which makes this way of correction
                less attractive);

                2) the integrated matrix element
        $ | M |^2 $
                is usually smooth enough (at least, piecewise)
                function of the variables and it might be
                well ap\-pro\-xi\-ma\-ted by a polynomial.

                Therefore,
                calculating the phase space integral
                in question,
                one deals with a classical case
                for which the Gauss integration method
                proved its unprecedented effectiveness.

                Let us note that there might be different
                approaches of implementation of
                Gauss method
                to the considered integral.
                One can try to find
                the set of orthogonal functions
                or polynomials of all four variables
                appropriate for the
                4--dimensional integral in question.
                This is very interesting and yet
                unsolved problem.
                Its solution should help much
                especially in the case
                when the integration must be performed
                over all the phase space
                (for example,
                when analyzing unitarity relation for the
                considered amplitude).
                However, when the integration is to be performed
                only over a part (bin)
                of the phase space
                or over a sequence of bins
                one easily finds that the
                set of functions providing
                calculations by the
                Gauss method
                must be distinct for every bin.
                This makes impossible to use once precalculated
                roots and corresponding weights of the
                polynomial system of the method.

                Here, we shall follow the way which can allow,
                in principle,
                to deal with all kinds of distributions in
                invariant variables: 1--, 2--, 3-- and
                4--dimensional.
                The separate treatment of all four integrals
                will be found to require only two types
                of orthogonal polynomials in every variable
                to be considered
                in the most general case.

\newpage

\subsection{ Standard  singularities of iterated integrals }

                To find the polynomial basis
                which is most suitable
                for the integral in question
                one must analyse the properties
                of all four iterated integrals
                in expressions
(\ref{rbst}),
(\ref{rAst}).

                Since it is the phase space
                induced specifics of the considered
                integral which is of interest
                we can take the amplitude in the
                expressions
(\ref{rbst}),
(\ref{rAst})
                as smooth as we like,
                say,
                polynomial.

                The singularity structure of
                the most interior integral over
        $ t_b $
                in eq.
(\ref{rbst})
                is evident.
                The theta--function cuts
                the entire interval between the roots
        ($ t_b^l $, $ t_b^r $)
                of
        $ D_4 $
                in this variable.
                In the case of a polynomial matrix element
        $ |M|^2 $
                this integral is an elementary one
                and it might be calculated analytically for
                every monomial term
\begin{equation}
\label{In}
      I_n \equiv
                \int_{t_b^l}^{t_b^r}
                \frac {t_b^n d t_b}
                      {\sqrt{ - D_4} }
            =
                \frac { 1}
                      {\sqrt{ - b_{t2}} }
                \int_{t_b^l}^{t_b^r}
                \frac {t_b^n d t_b}
                 {\sqrt{ ( t_b - t_b^l ) ( t_b^r - t_b )} }
\end{equation}
                by recurrent relation
\begin{equation}
\label{Inrec}
      I_n =
            -   \frac {2n -1}
                       { n }
                \frac { b_{t1}}
                      { 2 b_{t2}}
                    I_{n - 1}
            -   \frac {n -1}
                       { n }
                \frac { b_{t0}}
                      { b_{t2}}
                    I_{n - 2}
                \; .
\end{equation}
                This relation uniquely
                defines
        $ I_n $
                in terms of coefficients
        $ b_{t0} $, $ b_{t1} $, $ b_{t2} $
                of the single--variable form
(\ref{D4bt})
                of
        $ D_4 $.
                The values of the two starting members
                of the recurrent sequence
\begin{equation}
\label{I0I1}
      I_0 =
               \frac { \pi }
                      { \sqrt{ - b_{t2} } }
\; ; \;
                    I_1 =
               - \frac { b_{t1}}
                      { 2 b_{t2}}
                    I_{0}
                \;
\end{equation}
                show that
                integrand of eq.
(\ref{rbst}),
                being
                polynomial in
        $ s_b $,
                remains polynomial after first integration
                since the quantity
        $ b_{t2} $,
                entering
                both the square root and denominator
                of expressions
(\ref{Inrec}), (\ref{I0I1}),
                does not depend on
        $ s_b $.

                This leads to the important conclusions,
                namely:

                1. The natural weight function for the
                integral over
        $ t_b $
                is given by
\begin{equation}
\label{tbW}
                \mu_b ( t_b ) =
                \left[
                \sqrt{ ( t_b - t_b^l ) ( t_b^r - t_b )}
                \right]^{-1}
                        \; .
\end{equation}

                2. There is no nontrivial weight function
                for the next integration over
        $ s_b $.

                It is to be noted
                that the above conclusions are determined
                by the integration order chosen.
                The inverted conclusions will be made
                if one chooses the order of eq.
(\ref{rbts}):
                it is the integral over
        $ s_b $
                which acquires the weight function,
                originating from square root of
        $ D_4 $,
                the second integral in
        $ t_b $
                being free from nontrivial weight function.

                In the course of integration via relations
(\ref{In}),
(\ref{I0I1})
                the irrationality
\begin{equation}
\label{irrbt2}
                 \sqrt{ - b_{t2} }
                        \;
\end{equation}
                appears.
                It depends on
        $ s_a $, $ t_a $
                variables and might
                be imagined to be important for analysis
                of the subsequent integrations.
                In fact, it must be modified by the following
                integration over
        $ s_b $
                and the final answer appears to be more
                symmetric in terms of components
                of the elliptic form
(\ref{D4m}).

                Let us shift the
        $ s_b $, $ t_b $
                variables to make the answer
                less immense.
                The coordinates
        $ s_b^c $, $ t_b^c $
                of the ellipse center do depend on the
        $ s_a , t_a $
                variables.
                Fortunately, they were found to be smooth
                in the physical domain (see eqs.
(\ref{sbc})).
                The standard monomial in both variables
                is then
\begin{equation}
\label{M2bst}
      |M|^2  = M_{nm} \equiv (s_b - s_b^c)^{n} (t_b - t_b^c)^{m}
                \; .
\end{equation}

                By means of elementary calculations
                the details of
                which it is reasonable to omit,
                the integral
\begin{equation}
\label{rbnm}
   r_{nm} \equiv
   r_b (M_{nm})
                \;
\end{equation}
                can be brought to the form
                (provided
        $ n + m $
                is even;
                otherwise integral
        $ r_{nm} $
                is zero):
\begin{eqnarray}
\label{rnmst}
   r_{nm}   & = &
                \frac  {2\pi}
                      {(n + m + 1)!!}
                        \sqrt{\frac{b_c}
                                   {{\rm Det} \; \hat{b}}
                             }
                \left ( \frac {b_c}
                           {{\rm Det} \; \hat{b}}
                \right )^{\frac{n+m}
                                {2}}
                (-b_{02})^{\frac{n-m}
                                {2}}
                \times
\\
\nonumber
  & &
                \times
                \sum_{l=1}^{\left[\frac{m}{2}\right]}
                C_m^{2l} (n+m-2l-1)!!(2l-1)!!
                ({\rm Det} \; \hat{b})^l
                \left[b_{11}/2\right]^{m-2l}
                \; ,
\end{eqnarray}
                or, changing the order of integration, to the
                equivalent form:
\begin{eqnarray}
\label{rnmts}
   r_{nm}   & = &
                \frac  {2\pi}
                      {(n + m + 1)!!}
                        \sqrt{\frac{b_c}
                                   {{\rm Det} \; \hat{b}}
                             }
                \left ( \frac {b_c}
                           {{\rm Det} \; \hat{b}}
                \right )^{\frac{n+m}
                                {2}}
                (-b_{20})^{\frac{m-n}
                                {2}}
                \times
\\
\nonumber
  & &
                \times
                \sum_{l=1}^{\left[\frac{n}{2}\right]}
                C_n^{2l} (n+m-2l-1)!!(2l-1)!!
                ({\rm Det} \; \hat{b})^l
                \left[b_{11}/2\right]^{n-2l}
                \; .
\end{eqnarray}
                At first glance the quantity
        $ b_{02} $
        ($ b_{20} $)
                appears in the denominator of eq.
(\ref{rnmst})((\ref{rnmts}))
                when
        $ m > n $
        ($ n > m $).
                Then eq.
(\ref{rnmts})((\ref{rnmst}))
                helps to avoid the algebraic proof that
                the quantity in the denominator
                is exactly cancelled
                by the factor from the sum.

                Taking into account that coefficients
        $ b_{\alpha \beta} $,
                entering the answer,
                are (second order) polynomials of the
        $ s_a $, $ t_a $
                variables,
                one easily continues analysis of properties
                of the iterated integrals:

                3. The
                true
                irrationality generated by integration
                over
        $ s_b $, $ t_b $
                variables is
\begin{equation}
\label{irrbcdtb}
                        \sqrt{\frac{b_c}
                                   {{\rm Det} \; \hat{b}}
                             }
                =
                      2  \frac{ \sqrt{
       \left [
            s_a - ( m_1 + m_2)^2
       \right ]
       \left [
            s_a - ( m_1 - m_2)^2
       \right ]
                                     }
                              }
                       { s_a }
                \; .
\end{equation}

                4. Polynomial matrix element in
        $ t_a $
                variable remains polynomial after the
                considered internal integration;
                there is no nontrivial weight function
                for the next integration in
        $ t_a $
                variable.

                5. The discussed in the previous section
                asymmetry of the description
                of the phase space in the
                variables
        $ s_a $, $ t_a $
                appears to be deeper after the internal
                integration;
                the generic polynomial of
        $ s_a $
                variable acquires irrationality with the
                critical point just at the boundary of
        ($ s_a $, $ t_a $)--plot
\begin{equation}
            s_a = ( m_1 + m_2)^2 = s_a^0
                \;
\end{equation}
                multiplied by a rational function of
        $ s_a $.

                6. The position of the only pole
                of this rational function
\begin{equation}
            s_a = 0
                \;
\end{equation}
                is distant from the physical region,
                an expansion
                or polynomial ap\-pro\-xi\-ma\-ti\-on
                being allowed
                (in the potentially dangerous case of
                vanishing masses
        $ m_1 $, $ m_2 $
                both the pole and the irrationality
                disappear --- see eq.
(\ref{irrbcdtb})).

                Turning to the integration over
        $ t_a $
                variable one should remind
                that integration in eq.
(\ref{rAst})
                is assumed to be performed over
                domain cut out from the
        ($ s_a , t_a $)--plot
                by rectangle defined by bounds
        $ s_a^{j-1}, s_a^{j} $,
        $ t_a^{k-1}, t_a^{k} $
                of bin
        ($ j , k $).
                Whenever the rectangle is exactly
                inside the Chew--Low plot
                integration of any
        $ t_a $
                polynomial
                will provide only constant (in
        $ s_a $)
                contribution.
                If there is an interval in
        $ t_a $
                for which an arc of hyperbola
        $ D_{3a} = 0 $
                enters the integration domain as a part
                of boundary
                the
        $ t_a $
                integration results in an expression
                containing the term
        $ \sqrt{ A_t} $
                from the boundary function
(\ref{tagh}).
                The expression
(\ref{At})
                for discriminant
        $ A_t $
                shows that it is the absolute bound
        $ s_a^U = ( \sqrt{ s } - m_3 )^2 $
                where the analyticity might be lost
                due to the above square root
                --- another critical point
        $ s_a^H = ( \sqrt{ s } + m_3 )^2 $
                is well outside the physical region.
                (And again,
                the condition
        $  m_3 = 0 $
                bringing
        $ s_a^H $
                to the boundary
                simultaneously kills the square root
                itself.)

                Depending on the given binning in the
        $ s_a $, $ t_a $
                variables,
                the absolute bounds
        $ s_a^0 $, $ s_a^U $
                might be unattainable at all,
                only one might be attained and,
                at last, for a specific bin of a
        $ t_a $
                distribution in the case of some
                energy and particle masses
                both bounds might happen to undergo
                due counting.
                The bounds must always be
                taken into account
                when the total cross section
                is being calculated.

                Combining with the point 5. of
                the previous conclusion one can find
                that the last integration in
(\ref{rAst})
                must be performed either

                a) with the two--point singular expression
\begin{equation}
\label{irrsa0saU}
                        \sqrt{
       \left [
            s_a - ( m_1 + m_2)^2
       \right ]
       \left [
            ( \sqrt{ s } - m_3)^2 - s_a
       \right ]
                              }
                \; ,
\end{equation}
                or

                b) with
\begin{equation}
\label{irrsa0}
                        \sqrt{
       \left [
            s_a - ( m_1 + m_2)^2
       \right ]
                              }
                \; ,
\end{equation}
                or

                c) with
\begin{equation}
\label{irrsaU}
                        \sqrt{
       \left [
            ( \sqrt{ s } - m_3)^2 - s_a
       \right ]
                              }
                \; ,
\end{equation}
                or

                d) without square root singularity.

                Now, it is easy to realize
                that cases b) and c) admit a simple
                change of variables
\begin{equation}
\label{sa0ch}
            s_a = s_a^0 + x^2
                \; ,
\end{equation}
\begin{equation}
\label{saUch}
            s_a = s_a^U - x^2
                \; ,
\end{equation}
                respectively,
                which helps to get rid of singularity.
                Of course, the price is the doubling of
                the effective power of integrand
                (i.e. the power of
                ap\-pro\-xi\-ma\-ting polynomial for the smooth
                factor of integrand).
                This looks quite acceptable.
                In contrast, the case a) needs
                a nonpolynomial
                (trigonometric) substitution providing
                no guarantee for estimated rate
                of convergence.

                Finally, we can state that

                7. Depending on the given binning
                the last integration has the
                weight function given by eq.
(\ref{irrsa0saU}),
                namely,
\begin{equation}
\label{saW}
                \mu_A ( s_a ) =
                        \sqrt{
       \left (
            s_a - s_a^0
       \right )
       \left (
            s_a^U - s_a
       \right )
                              }
                \; ,
\end{equation}
                or might be processed with the trivial
                weight function;
                in the latter case
                a change of variable might be required.

                To resume the current subsection
                one should reread
                the discussed above points 1., ..., 7.
                and note that there is no principal
                difference between the weight function
        $ \mu_b ( t_b ) $
                of eq.
(\ref{tbW})
                and
        $ \mu_A ( s_a ) $.
                Indeed, by multiplication of numerator and
                denumerator of integrand by
        $ \mu_A ( s_a ) $
                one increases the effective power
                of the latter by one and converts the
                irrationality to the form of
        $ \mu_b $.
                Therefore, only {\it two} distinct sets
                of orthogonal polynomials
                are required for implementation
                of the Gauss method to the integration
                over the phase space in question, namely,
                the set, corresponding to the
                (normalized
                to the unit interval)
                weight function
        $ \mu_b $
                of eq.
(\ref{tbW}),
                and the set,
                corresponding to the unit weight function.

\newpage

\section { Principal Features of FORTRAN Code }

\subsection{
                Outlines of the Gauss method;
                specifics of realization
           }

                To proceed with implementation of results
                of the previous analysis
                let us briefly recall the basic formulae
                of the Gauss method
\cite{Gauss1866}.
                More details might be found
                in almost any book
                on numeric integrating.
                For example, see
\cite{StroudS66}.

                Suppose we have to calculate integral of the type
\begin{equation}
	\int_{-1}^1 f(x) \mu (x) dx
                \; ,
\end{equation}
                where
        $ f(x) $
                is a smooth enough function and
        $ \mu $ ---
                positive weight function which might have
                integrable singularities at the points
        $ \pm 1 $.
                Such weight function uniquely defines
                the set of polynomials
        $ \{\phi_n\} $,
        ($ {n \geq 0} $)
                orthogonal to each other with this weight function
                provided a normalization convention is adopted.
                Then the Gauss formula reads
\begin{equation}
\label{Gf}
        \int_{-1}^1 f(x) \mu (x) dx
                \simeq
                       \sum_{k=1}^n w_k^{(n)} f(x_k^{(n)})
         \equiv G_n
        \; .
\label{II}
\end{equation}
                Here, the {\it weights}
        $ \{w_k^{(n)}\} $
		and points
        $ \{x_k^{(n)} \in (-1,1)\} $
                are assumed to be chosen in such a way that
                relation
(\ref{II})
                turns into equality for any
        $ f $
                being a linear
                combination of the first
        $ n+1 $
                functions
        $ \phi_m $
		from the system
        $ \{\phi_m\} $.
                In this case
        $ x_k^{(n)} $ should be the roots of the
                polynomial
        $ \phi_{n} $.

                The function
        $ f $
                being a smooth function possesses
                rapidly decreasing Fourier
                co\-ef\-fi\-ci\-ents
        $ c_n $:
\begin{eqnarray}
   f(x) & = & \sum_{n\geq0} c_n \phi_n (x)
        \; ;
\\
   c_m & = & \int_{-1}^1 f(x) \phi_m \mu (x) dx
       \; .
\end{eqnarray}
                The convergence rate of the integral sums
        $ \{G_n\} $
		depends on that of
        $ \{c_n\} $
                and,
                con\-se\-quen\-t\-ly,
                on the choice of the system
	$ \{\phi_n\} $
                --- this
                opens the possibility to hide singularities
                of an overall integrand into the weight
                function
        $ \mu (x) $.

                According to the conclusion
                of the previous section
                the two cases of weight functions
                are peculiar to the iterated integrals over
                phase space, namely,
\begin{equation}
\label{mu}
              \mu ( x ) = (1-x^2)^{-1/2}
\end{equation}
                and
\begin{equation}
\label{mu0}
              \mu_0 ( x ) = 1
                \; .
\end{equation}

                In the first case the orthogonal functions
                are the Chebyshev polynomials
\begin{equation}
\label{Tn}
               T_n (x) = \cos{ ( n \, \arccos {x} ) }
                \; ,
\end{equation}
                for which
                there are exact expressions for
		roots and weights:
\begin{equation}
\label{xk}
               x_k^{(n)} = \cos{\pi\frac{2k-1}{2n}}
                \; ;
\end{equation}
\begin{equation}
\label{wk}
               w_k^{(n)} = \frac{\pi}{n}
                \; ,
\end{equation}
                the quadrature formula
(\ref{Gf})
                being
\begin{equation}
\label{GnT}
  \int_{-1}^1 f(x) (1-x^2)^{-1/2} dx \simeq
  \sum_{k=1}^n \frac{\pi}{n} f
                \left( \cos \left(\pi\frac{2k-1}{2n}\right)
                \right)
                \; .
\end{equation}

                The unit weight function defines the set
                of the well--known Legendre polynomials
\begin{equation}
\label{Pn}
         P_n(x) = \frac{1}{2^nn!}\frac{d^n}{dx^n}(1-x^2)^n
                \; .
\end{equation}
                In this case the roots
        $ X_k^{(n)} $
                of the Legendre polynomial
        $ P_{n} $
                for
        $ n \geq 10 $
                are to be found only by numerical solution
                of algebraic equation
        $ P_{n} (x) = 0 $.
                The corresponding weights
        $ W_k^{(n)} $
                are then given by the expression
\begin{equation}
\label{Wk}
               W_k^{(n)} =
  2 [1- (X_k^{(n)})^2]^{-1}
        \left[
\frac{d}{dx} P_n(x) \vert_{x=X_k^{(n)}}
        \right]^{-2}
                \; .
\end{equation}
                Explicit expression for the considered
                integral is
\begin{equation}
\label{GnP}
  \int_{-1}^1 f(x) dx \simeq
  2 \sum_{k=1}^n [1- (X_k^{(n)})^2]^{-1}
        \left[
\frac{d}{dx} P_n(x) \vert_{x=X_k^{(n)}}
        \right]^{-2} f(X_k^{(n)})
                \; .
\end{equation}

                The formulae
(\ref{GnT}),
(\ref{GnP})
                are the basis of the central subroutines
                of the integrating program,
                namely, {\sf CHIN} and {\sf ZNIN}.
                To avoid problems with recurrent calls
                in the course of processing iterated
                integrals the subroutines were simply
                cloned: identical copies
                {\sf CHIN1}, {\sf CHIN4}
                and
                {\sf ZHIN1}, {\sf ZHIN2}, {\sf ZHIN3}
                were used to be called
                at the corresponding level of integration.

                Whereas the programming of
                a subroutine of the {\sf CHIN} type
                (based on Chebyshev polynomials)
                is straightforward
                {\sf ZHIN} subroutines require
                calculation of roots of Le\-gen\-dre
                polynomials
                and the corresponding weights in
                much more complicated man\-ner.
                The\-re\-fo\-re,
                to
                avoid significant losses of time during
                integration runs
                the way of using precalculated values
                was chosen.

                The roots of Legendre polynomials
                have the following property:
                any root
                of
                every subsequent polynomial is located
                between corresponding roots of the previous
                Legendre polynomial.
                Hence,
                the Newton method of determination of
                roots is
                the most suitable in this case.
                It operates with function and its
                derivative.
                Legendre polynomials
        $ P_l $
                and the derivatives
        $ D_l $
                might be
                easily calculated by  the
                well--known recurrent formulae:
\begin{equation}
    l P_l (x)  =  x P_{l-1}(x) (2l - 1)  -  P_{l-2}(x) (l-1)
                \; ;
\end{equation}
\begin{equation}
\label{Dl}
     D_l (x)  =  (2l - 1) P_{l-1}(x)  +  D_{l-2}(x)
                \; ,
\end{equation}
                starting the recurrence from
        $ P_0 = 1 $,
        $ P_1 = x $
                and
        $ D_0 = 0 $,
        $ D_1 = 1 $.

                As soon as the roots are found
                the values of corresponding weights
                are determined by eq.
(\ref{Wk})
                in terms of roots and derivatives
(\ref{Dl}).

                Since the above calculations are needed
                to be performed only once
                the time of run is not of much importance
                but the precision is,
                because it directly determines the
                accuracy of the integration method.
                To avoid complicated analysis
                of the influence of the uncertainties
                of roots and weights on the final answer
                the latter were calculated
                at the maximal precision allowed by
                VAX/VMS PASCAL.
                The present realization contains
                common block {\sf RTWT}
                with roots and weights
                for Legendre polynomials up to 50 order
                (up to 100 --- available);
                in practice, the maximal order 16
                appears to be called.

\newpage
\subsection{
                Implementation
           }

                The principle structure of the program is
                simple:
                User defined
                {\sf MAIN}
                manipulates with User data base to provide
                necessary parameters,
                calls integration subroutine
                {\sf WTFF}
                and performs appropriate output of results.
                There must be another User routine which is
                called by
                {\sf WTFF},
                namely, matrix element
                {\sf USRF}.


                The numeric integration
                over 4--dimensional phase space
                is performed by
                subroutine {\sf WTFF}.
                It returns the value of integral and
                its absolute difference with the value of
                previous iteration which are the variables
                of the argument list.
                The list of physical variables to be passed
                to this procedure includes energy
        $ s $,
                particle masses
        $ m_a $, $ m_b $, $ m_1 $, $ m_2 $, $ m_3 $
                and bounds of 2--dimensional bin
        $ s_a^{j-1}, s_a^{j} $,
        $ t_a^{k-1}, t_a^{k} $.
                Only the bounds were chosen to be included
                into the list of arguments --- other
                variables are accessible
                through the {\sf COMMON}
                blocks.
                This is because of the following reasons:
                a) along with the value of
        $ s $
                some other totally equivalent
                precalculated quantities
        $ T_{\rm BEAM} $,
        $ E_{\rm BEAM} $,
        $ P_{\rm BEAM} $,
                etc. are necessary also for calculation of
                matrix element provided by User;
                b) according to the described in the end of
                section 2 universal way of treating bins in
                one or another variable the above list of
                masses realizes some transposition of particle
                masses of the User reaction; a manipulation is
                needed to pass true arguments to the amplitude
                which should not know anything about this.

                Apart the above physical variables the
                subroutine {\sf WTFF}
                operates with a number of
                control parameters including terminating
                value EPS of attained accuracy and 4 limits of
                iterations for all 4 integrals.

                The {\sf WTFF} algorithm
                exploits the crucial property of phase
                space which is expressed by the fact that at
                any given values of variables
        $ s_a $
                and
        $ t_a $
                the allowed domain of variables
        $ s_b $, $ t_b $
                is the ellipse.
                Parameters of the latter are
                determined by masses of external particles and
                the values of
        $ s_a $
                and
        $ t_a $
                --- the corresponding functions and
                subroutines are built according to the
                formulae given in section 3 and in Appendix.

                The integration domain in the plain
          $ ( s_a, t_a ) $
                is the intersection of the given
                rectangle
\begin{eqnarray}
\label{rect}
         s_a^{j-1} \leq s_a \leq s_a^{j} \; ; \\
\nonumber
         t_a^{k-1} \leq t_a \leq t_a^{k} \;
\end{eqnarray}
                with the region
\begin{eqnarray}
\label{satargn}
         s_a^{D} \leq s_a \leq s_a^{U} \; ; \\
\nonumber
         t_a^{g}(s_a) \leq t_a \leq t_a^{h}(s_a) \;
\end{eqnarray}
                (see eqs.
(\ref{sarange}), (\ref{tagh})).

                The properties of the resulting domain
                determine the integration method
                (based on Legendre or Chebyshev polynomials)
                 to be applied to the
        $ s_a $
                variable.
                In the most general case we have to split
                exterior integration into up to 4
                processes (intervals)
                ({\sf SINTA}, ..., {\sf SINTD})
                so that
                boundaries
                of the internal integration over
        $ t_a $
                have smooth dependence on
        $ s_a $
                on each interval.
                Thus every process performs
                integration of the
                smooth function
         {\sf FWA}, {\sf FWB}, {\sf FWC} or {\sf FWD}
                (which are specified below)
                and admits an exact method.

                As it is discussed in
                subsection 4.2 almost each of the four
                intervals might contain singularities
                at
                their ends
                classified by the cases
                a), b), c) and d).
                So,
                depending on the relative values of
                integration interval bounds and absolute
                bounds
(\ref{sarange}),
                the calculation of
                each of integrals
                {\sf SINTA}, ..., {\sf SINTD}
                is processed by
                different algorithms,
                using procedures
                {\sf CHIN1}
                or {\sf ZHIN1}.
                The algorithms call the corresponding
                functions
                {\sf FWA}, {\sf FWB}, {\sf FWC} or {\sf FWD}
                depending on the case of singularities
                classified by
                a), b), c), d).
                These functions realize appropriate changes of
                variable
        $ s_a $
                and make this problem hidden for the
                next integration over variable
        $ t_a $.

                The analysis of the previous section
                (see points 1., ..., 7.)
                shows that
                the next two integrations over
                any bin in variable
         $ t_a $
                and over
                the complete range
         ($ s_b^L $, $ s_b^R $)
                of
         $ s_b $
                variable
                are performed
                by Gauss method with Legendre polynomials.
                The algorithms in question are realized by
                identical subroutines
                {\sf ZHIN2} and {\sf ZHIN3}.

                According to the quoted above analysis
                the last integration is processed by
                sub\-rou\-tine
                {\sf CHIN4}
                which realizes a simple
                algorithm of Gauss integration
                with Che\-by\-shev po\-ly\-no\-mials
                over
                unnormalized interval
        ($ t_b^{l} $, $ t_b^{r} $)
                given by eq.
(\ref{tblr}).
                The call of the User matrix element
                {\sf USRF} of
         ($ s_a, t_a,  s_b, t_b $)
                arguments arises at this stage.
                Prior to the call the initialization of all
                ten scalar products
(\ref{pvar}),
(\ref{avar})
                is performed.
                The scalar products
                (like the particle masses)
                are available via
                {\sf COMMON}
                block.
                This must simplify the calculation of the
                considered amplitude
                which is usually derived according to
                Feynman rules.
                However, User must provide appropriate
                rearrangement of the scalar products
                if variables of the diagram
(1)
                are a permutation of the User ones.

                In total, the discussed program
                contains about four dozens of functions
                and sub\-rou\-ti\-nes
                handling various kinematic calculations.
                The detailed description will follow
                the specifics of the phase space geometry
                discussed in Sect. 3 and collection of
                formulae given in Appendix.

\newpage
\section {
                Conclusion
         }

                In the result of analysis of the geometry
                of relativistic 3--particle phase space
                and implementation of integration algorithm
                the principal
                solvability of the integration problem
                appears to be  demonstrated.
                Besides,
                it becomes evident that the considered case
                of
    2 $ \rightarrow $ 3
                reaction
                is likely to be the last one permitting
                a treatment outside the framework of
                Monte Carlo approach.
                This is due to combinatorical growth of
                the number of kinematical variables
                (as free as well as dependent ones)
                characterizing the reaction
                (as compared with
    2 $ \rightarrow $ 2
                case)
                and the complication of geometry
                which is displayed by growing
                algebraic power of definition equations.
                Given more than 3 particles in the
                final state
                other methods of analysis,
                like the artificial intelligence ones,
                become necessary.

                The integration problem has been solved
                here for 1--dimensional distributions
                in any of
        $ 10 - 1 = 9 $
                two--particle invariant variables
                (i.e. without counting initial energy
        $ s $)
                of the considered reaction
                on the base of
                the fast integration algorithm
                implementing the Gauss method.
                The algorithm needs to process the
                calculation of matrix element in few
                thousands of points in phase space
                (compared with
                tens of millions ones and even more in the
                Monte Carlo routines)
                and provides the accuracy which
                will be never attainable by
                Monte Carlo calculations.

                It goes without saying
                that the solution was possible due to
                the extensive coverage of
                the
    2 $ \rightarrow $ 3
                case by preceding
                investigations
                (the reader had been already advised
                to look the books
\cite{BK,Kopylov70}
                for an overview).
                The same reason might be used
                for motivation of further researches
                in this field.
                At least two directions of the researches
                might be proposed:
                1) a generalization of the approach for
                the case of noninvariant variables
                (like angles and 3--momenta);
                2) its extension to treat
                2--, 3-- and 4--dimensional distributions.

                While the first direction seems to be
                closer to the needs of specific
                experiments
                the second one appears to be more attractive.
                Indeed, now it is not so difficult
                to form almost any distribution
                (hence, in invariant variables as well)
                from full--kinematics data with the help
                of modern tools like GEANT--3.
                The capability of comparison of different
                measurements will be opened then
                and the phenomenological parameters
                describing specifics of distributions
                will get an universal status.

                The field of application
                of the considered integration algorithm
                might include a large variety of processes
                at intermediate energies like
        $  \pi D \rightarrow \pi p n $,
        $  \pi N \rightarrow \pi \pi N $,
        $  \gamma p  \rightarrow \pi \pi N $,
                etc.
                One can also try the approach
                at high energies as well
                (at least for control of accuracy of
                the ultra--relativistic approximation for
                phase space).

\section
        { Acknowledgments }

                We thank CHAOS team and DEC
                for providing resources of ALPHA processors
                at TRIUMF (Vancouver), INFN (Trieste) and
                AXP/OSF (Pasadena).
                AAB is grateful to
                colleagues of CHAOS team for interest,
                hospitality and help during his stay at Canada
                and Italy.

                This research was supported in part
                by RFBR grant N 95-02-05574a.

\newpage

\vskip 1cm

\newpage

\setcounter{equation}{0}
\renewcommand{\theequation}{A.\arabic{equation}}

\section{ Appendix: Gram Determinants }
\subsection{ Two--variable matrix form of $ D_4 $ }

                The fourth order Gram determinant
        $ D_4 $
                does not depend on the choice of
                particle momenta.
                Let us define it as follows:
\begin{eqnarray}
\nonumber
             D_4 & \equiv & - \Delta_4 ( q_2 , q_3 , p_b , p_a )
\\
  & = &       -
                 \left|
\begin{array}{cccc}
                  { q_2 \cdot q_2 } &
                     \underline{ q_2 \cdot q_3 } &
                        \underline{ q_2 \cdot p_b } &
                           \underline{ q_2 \cdot p_a } \\
                  \underline{ q_3 \cdot q_2 } &
                     { q_3 \cdot q_3 } &
                        { q_3 \cdot p_b } &
                           { q_3 \cdot p_a } \\
                  \underline{ p_b \cdot q_2 } &
                     { p_b \cdot q_3 } &
                        { p_b \cdot p_b } &
                           { p_b \cdot p_a } \\
                  \underline{ p_a \cdot q_2 } &
                     { p_a \cdot q_3 } &
                        { p_a \cdot p_b } &
                           { p_a \cdot p_a }
\end{array}
                             \right|
                  \; .
\end{eqnarray}
                Here,
                all scalar products depending on
        $ s_b $
                and
        $ t_b $
                variables are underlined
                (see eqs.
(\ref{pvar}),
(\ref{avar})).
                This makes evident that
        $ D_4 $
                is only quadratic in
        $ s_b $
                and
        $ t_b $.
                By virtue of the
                symmetry
        ($ a \leftrightarrow b $,
        $ 1 \leftrightarrow 3 $)
                the same conclusion is valid for
        ($ s_a $, $ t_a $)
                pair as well.

                Expression for
        $ D_4 $
                in terms of expansion in
        $ s_a, t_a, s_b, t_b $:
\begin{eqnarray}
\label{dxxxx}
 D_4 & = & \sum_{\alpha_s,\alpha_t,\beta_s,\beta_t}
               d_{\alpha_s,\alpha_t,\beta_s,\beta_t}
      s_a^{\alpha_s} t_a^{\alpha_t} s_b^{\beta_t} t_b^{\beta_t} \;
\\
\label{Appbxx}
 & = & \sum_{\beta_s,\beta_t}
               b_{\beta_s,\beta_t}
                                    s_b^{\beta_t} t_b^{\beta_t} \;
\end{eqnarray}
                is determined by nonzero coefficients
\begin{eqnarray}
\nonumber
d_{0 0 0 0} & = & ( - (2 (((m_b^2 + m_a^2 - s) m_3^2 + m_b^2 m_a^2
 - m_a^4 + m_a^2 s) m_2^2  - m_3^4 m_b^2
\\
 & &
\nonumber
 + m_3^2 m_b^2 m_a^2) m_1^2
+ 2 (((m_a^2 + s) - m_b^2) m_3^2
 - 2 m_a^2 s) m_2^2 m_b^2
\\
 & &
\nonumber
+ (m_3 + m_a)^2 (m_3 - m_a)^2 m_1^4
\\
 & &
\nonumber
+ (m_b^2 + 2 m_b m_a + m_a^2 - s) (m_b^2 - 2 m_b m_a
+ m_a^2 - s) m_2^4
\\
 & &
\nonumber
   + m_3^4 m_b^4))
/16 \; ; \\
\nonumber
d_{0 0 0 1} & = & (((2 m_b^2 - m_a^2 - s) m_3^2
+ m_3^4 + m_a^2 s) m_1^2
\\
 & &
\nonumber
- ((m_b^2 - m_a^2 + s) m_3^2
+ m_b^2 s + m_a^2 s - s^2) m_2^2
\\
 & &
\nonumber
   + m_3^4 m_b^2 - m_3^2 m_b^2 s)/8 \; ; \\
\nonumber
d_{0 0 0 2} & = & ( - (m_3^2 - s)^2)/16 \; ; \\
\nonumber
d_{0 0 1 0} & = & ( - (((m_a^2 + s) - 2 m_3^2 - m_b^2) m_2^2
\\
 & &
\nonumber
+ (m_3 + m_a) (m_3 - m_a) m_1^2 - m_3^2 m_b^2) m_b^2)/8 \; ; \\
\nonumber
d_{0 0 1 1} & = & ( - (m_3^2 - s) m_b^2)/8 \; ; \\
\nonumber
d_{0 0 2 0} & = & ( - m_b^4)/16 \; ; \\
\nonumber
d_{0 1 0 0} & = & (((m_b^2 - m_a^2 - s) m_2^2
- (m_b^2 - 2 m_a^2 + s) m_3^2 - m_a^2 s) m_1^2
\\
 & &
\nonumber
+ (m_3^2 + m_a^2) m_1^4
- (m_b^2 + m_a^2 - s) m_2^2 s + m_3^2 m_b^2 s)/8 \; ; \\
\nonumber
d_{0 1 0 1} & = & ( - ((m_3^2 + s) m_1^2 - 2 m_2^2 s + m_3^2 s -
s^2))/8 \; ; \\
\nonumber
d_{0 1 1 0} & = & ( - ((m_3^2 + m_b^2 + m_a^2 - 2 s) m_1^2
\\
 & &
\nonumber
+ (m_b^2 - m_a^2 + s) m_2^2 + m_3^2 m_b^2 + m_b^2 s))/8 \; ; \\
\nonumber
d_{0 1 1 1} & = & (m_3^2 - s)/8 \; ; \\
\nonumber
d_{0 1 2 0} & = & m_b^2/8 \; ; \\
\nonumber
d_{0 2 0 0} & = & ( - (m_1^2 - s)^2)/16 \; ; \\
\nonumber
d_{0 2 1 0} & = & (m_1^2 + s)/8 \; ; \\
\nonumber
d_{0 2 2 0} & = & ( - 1)/16 \; ; \\
\nonumber
d_{1 0 0 0} & = & (((2 m_2^2 - m_3^2 + m_a^2) m_1^2
- (m_b^2 - m_a^2 + s) m_2^2 + m_3^2 m_b^2) m_a^2)/8 \; ; \\
\nonumber
d_{1 0 0 1} & = & ( - ((m_3^2 + m_a^2) m_1^2
+ (m_b^2 + m_a^2 - 2 s) m_3^2
\\
 & &
\nonumber
- (m_b^2 - m_a^2 - s) m_2^2 + m_a^2 s))/8 \; ; \\
\nonumber
d_{1 0 0 2} & = & (m_3^2 + s)/8 \; ; \\
\nonumber
d_{1 0 1 0} & = & ((m_3 + m_a) (m_3 - m_a) m_1^2
\\
 & &
\nonumber
- (m_b^2 + m_a^2 - s) m_2^2 - m_3^2 m_b^2
- m_b^2 m_a^2)/8 \; ; \\
\nonumber
d_{1 0 1 1} & = & ( - (m_3^2 + m_b^2 - 2 m_a^2 + s))/8 \; ; \\
\nonumber
d_{1 0 2 0} & = & m_b^2/8 \; ; \\
\nonumber
d_{1 1 0 0} & = & ( - (m_1^2 - s) m_a^2)/8 \; ; \\
\nonumber
d_{1 1 0 1} & = & (m_1^2 - s)/8 \; ; \\
\nonumber
d_{1 1 1 0} & = & ( - (m_1^2 - 2 m_b^2 + m_a^2 + s))/8 \; ; \\
\nonumber
d_{1 1 1 1} & = & ( - 1)/8 \; ; \\
\nonumber
d_{1 1 2 0} & = & 1/8 \; ; \\
\nonumber
d_{2 0 0 0} & = & ( - m_a^4)/16 \; ; \\
\nonumber
d_{2 0 0 1} & = & m_a^2/8 \; ; \\
\nonumber
d_{2 0 0 2} & = & ( - 1)/16 \; ; \\
\nonumber
d_{2 0 1 0} & = & m_a^2/8 \; ; \\
\nonumber
d_{2 0 1 1} & = & 1/8 \; ; \\
\nonumber
d_{2 0 2 0} & = & ( - 1)/16 \; .
\end{eqnarray}
             Coefficients
        $  b_{\beta_s,\beta_t} $
             are built of combinations of the latter ones with
             cor\-res\-pon\-ding powers of
        $ (s_a, t_a) $:
\begin{eqnarray}
\nonumber
b_{0 0} & = &
           ( - (2 (((m_b^2 + m_a^2 - s) m_3^2
                  + (m_a^2 - t_a) m_b^2
\\
 & &
\nonumber
                  + (s - 2 s_a + t_a) m_a^2 - m_a^4 + s t_a) m_2^2
\\
 & &
\nonumber
                 + ((m_a^2 + t_a) m_b^2
                  + (s_a - 2 t_a) m_a^2 + s t_a) m_3^2
\\
 & &
\nonumber
                 + (s + s_a) m_a^2 t_a
                 - m_3^4 m_b^2 - m_a^4 s_a - s t_a^2) m_1^2
\\
 & &
\nonumber
              + 2 (((m_a^2 + s) - m_b^2) m_3^2 m_b^2
                 - ((2 s - s_a) m_a^2 - s t_a) m_b^2
\\
 & &
\nonumber
                 + (s_a + t_a) m_a^2 s - m_a^4 s_a - s^2 t_a) m_2^2
\\
 & &
\nonumber
                 + (m_3^2 + 2 m_3 m_a + m_a^2 - t_a)
                   (m_3^2 - 2 m_3 m_a + m_a^2 - t_a) m_1^4
\\
 & &
\nonumber
                 + (m_b^2 + 2 m_b m_a + m_a^2 - s)
                   (m_b^2 - 2 m_b m_a + m_a^2 - s) m_2^4
\\
 & &
\nonumber
               - 2 (m_a^2 s_a + s t_a) m_3^2 m_b^2
               + m_3^4 m_b^4 + m_a^4 s_a^2
               - 2 m_a^2 s s_a t_a + s^2 t_a^2))/16
 \; ; \\
\nonumber
b_{0 1} & = & (((2 m_b^2 - m_a^2 - s - s_a - t_a) m_3^2
              + (s - s_a) m_a^2 + m_3^4 - s t_a + s_a t_a) m_1^2
\\
 & &
\nonumber
       - ((m_b^2 - m_a^2 + s) m_3^2
+ (s + s_a) m_a^2 + (s - s_a) m_b^2
\\
 & &
\nonumber
+ (s_a - 2 t_a) s - s^2) m_2^2
\\
 & &
\nonumber
- ((s + s_a) m_b^2 - (2 s_a - t_a) s
  + m_a^2 s_a) m_3^2 - (s - s_a) m_a^2 s_a
\\
 & &
\nonumber
  + m_3^4 m_b^2 + s^2 t_a - s s_a t_a)/8 \; ; \\
\nonumber
b_{1 0} & = &
           ( - (((m_b^2 - s_a + t_a) m_3^2
               - (m_a^2 - t_a) m_b^2
\\
 & &
\nonumber
               + (s_a + t_a) m_a^2 - 2 s t_a
               + s_a t_a - t_a^2) m_1^2
\\
 & &
\nonumber
              + ((m_a^2 + s + s_a + t_a) m_b^2
               + (m_a^2 - s) (s_a - t_a)
                - 2 m_3^2 m_b^2 - m_b^4) m_2^2
\\
 & &
\nonumber
              + ((s_a + t_a) - m_b^2) m_3^2 m_b^2
               + (m_a^2 s_a + s t_a - 2 s_a t_a) m_b^2
\\
 & &
\nonumber
               - (m_a^2 s_a - s t_a) (s_a - t_a)))/8 \; ; \\
\nonumber
b_{1 1} & = & ( - (((s_a - t_a) + m_b^2) m_3^2
            - (s - s_a) m_b^2 + (s_a+ t_a) s
\\
 & &
\nonumber
             - (s_a - t_a) s_a - 2 m_a^2 s_a))/8 \; ; \\
\nonumber
b_{0 2} & = & (2 (s + s_a) m_3^2 - m_3^4
               - s^2 + 2 s s_a - s_a^2)/16
 \; ; \\
\label{bxx}
b_{2 0} & = & (2 (s_a + t_a) m_b^2 - m_b^4
               - s_a^2 + 2 s_a t_a - t_a^2)/16
 \; .
\end{eqnarray}

             When considered as form of
        $ (s_b, t_b) $
             variables
        $ D_4 $
             might be written as
\begin{equation}
\label{D4f2}
         D_4 =
                  \left(
\begin{array}{c}
                   s_b \\
                   t_b
\end{array}
                             \right)^{\rm T}
                \cdot \hat{b} \cdot
                 \left(
\begin{array}{c}
                   s_b \\
                   t_b
\end{array}
                             \right)
                + b^{\rm T}  \cdot
                 \left(
\begin{array}{c}
                   s_b \\
                   t_b
\end{array}
                             \right)
                + b_{00} \; ,
\end{equation}
               where
\begin{equation}
                 b \equiv
                 \left(
\begin{array}{c}
                   b_{10} \\
                   b_{01}
\end{array}
                             \right)
                  \; ; \;
                 \hat{b} \equiv
                 \left(
\begin{array}{cc}
                   b_{20} & b_{11}/2 \\
                   b_{11}/2 & b_{02}
\end{array}
                             \right)
                  \; .
\end{equation}

             Eliminating the linear terms in the form
(\ref{D4f2})
             by shift of variables
        $$
                 \left(
\begin{array}{c}
                   s_b \\
                   t_b
\end{array}
                             \right)
                  \; = \;
                 \left(
\begin{array}{c}
                   s_b' \\
                   t_b'
\end{array}
                             \right)
                  \;  +  \;
                 \left(
\begin{array}{c}
                   s_b^c \\
                   t_b^c
\end{array}
                             \right)
                                \; ,
        $$
                where
\begin{equation}
\label{sbc}
                 \left(
\begin{array}{c}
                   s_b^c \\
                   t_b^c
\end{array}
                             \right)
                  \;  =  \;
                - \frac{1}
                       {2}
                 \hat{b}^{-1} \cdot b
\end{equation}
        $$
                  \;  =  \;
                 \left(
\begin{array}{c}
        (
              - ( s - s_a - m_3^2 ) m_1^2
              + ( s + s_a - m_3^2 ) m_2^2
              + ( s - s_a + m_3^2 ) s_a
        )  / ( 2 s_a ) \\

        (
                ( s_a + t_a - m_b^2 ) m_1^2
              + ( s_a - t_a + m_b^2 ) m_2^2
              - ( s_a - t_a - m_b^2 ) s_a
        )  / ( 2 s_a )
\end{array}
                             \right)
                        \; ,
        $$
             it is possible to rewrite
        $ D_4 $
             as the {\it centered} form
\begin{equation}
\label{D4c}
         D_4 =
                  \left(
\begin{array}{c}
                   s_b' \\
                   t_b'
\end{array}
                             \right)^{\rm T}
                \cdot \hat{b} \cdot
                 \left(
\begin{array}{c}
                   s_b' \\
                   t_b'
\end{array}
                             \right)
                + b_{c} \; .
\end{equation}
             The free term
        $ b_c $
             of this form is
\begin{equation}
         b_c  =   b_{00}
                  \;  -  \;
                  {{1}\over{4}}
                 b^{\rm T} \cdot \hat{b}^{-1} \cdot b
                  \;  =  \;
           D_{2a}  D_{3a}
             /  s_a  \; .
\label{bc}
\end{equation}
             Here,
        $ D_{2a} $
             and
        $ D_{3a} $
             are the Gram determinants
\begin{equation}
\label{D2sf}
          D_{2a}  \equiv - \Delta_2 ( q_1 , q_2 )
                  \;  =  \;
       (
            s_a - ( m_1 + m_2)^2
       )
       (
            s_a - ( m_1 - m_2)^2
       )
                 / 4
            \; ,
\end{equation}
\begin{eqnarray}
\nonumber
          D_{3a} & \equiv & \Delta_3 ( q_3 , p_b , p_a )
\\
\nonumber
          & = &
      (
           - s_a m_a^4 - t_a s^2 + t_a s (s_a - t_a)
\\
\label{D3f}
 & &
          + ((m_a^2 + s + s_a + t_a) m_b^2
                + (m_a^2 - s) (s_a - t_a) - m_b^4) m_3^2
\\
 & &
\nonumber
          + ((s_a + t_a) s - s_a^2 + s_a t_a) m_a^2
             - (m_a^2 - t_a) (s - s_a) m_b^2 - m_3^4 m_b^2
       )
        /  4  \; .
\end{eqnarray}
             In its turn the determinant of the matrix
        $ \hat{b} $
             of the quadratic form
(\ref{D4f2})
             which remains the same for the centered case
(\ref{D4c})
             also contains the
        $ D_{3a} $
             factor:
\begin{eqnarray}
         {\rm Det} \; \hat{b}  =
          s_a
            D_{3a}
        / 16  \; .
\label{Dtb}
\end{eqnarray}

\subsection{ Single--variable form of $ D_4 $ }

                When considered as function
                of a single variable
                (because of symmetry
        $ a \leftrightarrow b $
                the variable is reasonable to choose
        $ s_b $ or $ t_b $)
        $ D_4 $
                is determined by expansions
\begin{equation}
\label{D4bs}
          D_4
                  \;  =  \;
                b_{s2} s_b^2 + b_{s1} s_b + b_{s0}
            \; ,
\end{equation}
\begin{equation}
\label{D4bt}
          D_4
                  \;  =  \;
                b_{t2} t_b^2 + b_{t1} t_b + b_{t0}
            \; ,
\end{equation}
                where
\begin{eqnarray}
\nonumber
         b_{s2} & = & b_{20} \; ; \\
\label{bs20}
         b_{s1} & = & b_{11} t_b + b_{10} \; ; \\
\nonumber
         b_{s0} & = & b_{02} t_b^2 + b_{01} t_b + b_{00}
                \;
\end{eqnarray}
                and
\begin{eqnarray}
\nonumber
         b_{t2} & = & b_{02} \; ; \\
\label{bt20}
         b_{t1} & = & b_{11} s_b + b_{01} \; ; \\
\nonumber
         b_{t0} & = & b_{20} s_b^2 + b_{10} s_b + b_{00}
                \;
\end{eqnarray}
                are defined by expressions
(\ref{bxx})
                for coefficients
        $  b_{\beta_s,\beta_t} $.

                To calculate the roots
                of the second order polynomials
(\ref{D4bs}), (\ref{D4bt})
                one needs in fact only
                coefficients
        $  b_{s2} $
                and
        $  b_{s1} $
        ($  b_{t2} $
                and
        $  b_{t1} $)
                provided the discriminants of
(\ref{D4bs}), (\ref{D4bt})
\begin{equation}
\label{dscbsf}
          b_{s}
                  \;  =  \;
                b_{s1}^2 - 4 b_{s2} b_{s0}
            \; ,
\end{equation}
\begin{equation}
\label{dscbtf}
          b_{t}
                  \;  =  \;
                b_{t1}^2 - 4 b_{t2} b_{t0}
            \;
\end{equation}
                are known.
                The relevant discussion of Jacobi
                reduction theorem for symmetric
                de\-ter\-mi\-nants
                of arbitrary dimension might
                be found, for example in
\cite{Morrow66,Tarski60}.
                In our particular case
                of the fourth order Gram determinant
        $ D_4 $
                the results might be verified by direct
                calculation to be the products of Gram
                determinants
\begin{equation}
\label{dscbs}
          b_{s}
                  \;  =  \;
                 D_{3a} D_{3as}
            \; ,
\end{equation}
\begin{equation}
\label{dscbt}
          b_{t}
                  \;  =  \;
                 D_{3a} D_{3at}
            \; ,
\end{equation}
                where explicit expression for
        $ D_{3a} $
                is given in
(\ref{D3f})
                and the ones for
        $ D_{3as} $, $ D_{3at} $
                are
\begin{eqnarray}
\nonumber
 D_{3as} & = &  \{
   ((m_b^2 - s_a + t_a) m_2^2 - (t_b - t_a) m_b^2
   + (s_a + t_a) t_b + s_a t_a - t_a^2) m_1^2
\\
\label{D3Bs}
& &
  + ((t_b + s_a + t_a) m_b^2 + (s_a - t_a) t_b - m_b^4) m_2^2
\\ \nonumber & &
  + (t_b - t_a) m_b^2 s_a
  - (s_a - t_a) t_b s_a - m_1^4 t_a - m_2^4 m_b^2 - t_b^2 s_a \}
        / 4 \; ;
\end{eqnarray}
\begin{eqnarray}
\nonumber
 D_{3at} & = &  \{
   (((m_3^2 + s - s_a) m_2^2
   + (s + s_a + s_b) m_3^2 - m_3^4 - s s_b + s_a s_b) m_1^2
\\
\label{D3Bt}
& &
  + ((s - s_b) m_3^2 + (s_a + s_b) s - s^2 + s_a s_b) m_2^2
\\ \nonumber & &
  - (s - s_b) m_3^2 s_a - m_1^4 m_3^2 - m_2^4 s
     + s s_a s_b - s_a^2 s_b - s_a s_b^2)\}
        / 4 \; .
\end{eqnarray}

                For processing the boundaries of
        ($ s_b $, $ t_b $)--plot
                at given fixed values of
        $ s_a $ and $ t_a $
                variables the similar treatment of
        $ D_{3as} $ and $ D_{3at} $ determinants is
                necessary.
                Expanding the latter in the form
\begin{equation}
\label{D3bst}
              D_{3as}     \;  =  \;
                b_{st2} t_b^2 + b_{st1} t_b + b_{st0}
            \; ;
\end{equation}
\begin{equation}
\label{D3bts}
              D_{3at}
                  \;  =  \;
                b_{ts2} s_b^2 + b_{ts1} s_b + b_{ts0}
            \; ,
\end{equation}
                where
\begin{eqnarray}
\nonumber
  b_{st0} & = &  \{
   ((m_b^2 - s_a + t_a) m_2^2
     + m_b^2 t_a + s_a t_a - t_a^2) m_1^2
\\
\label{bst0}
& &
  + ((s_a + t_a) - m_b^2) m_2^2 m_b^2 - m_1^4 t_a - m_2^4 m_b^2
     - m_b^2 s_a t_a)\}
          /4      \; ;
\\
\label{bst1}
  b_{st1} & = &  \{
   (m_b^2 + s_a - t_a) m_2^2
  - (m_b^2 - s_a - t_a) m_1^2
\\
\nonumber
& &
   + m_b^2 s_a - s_a^2 + s_a t_a \}
          /4      \; ;
\\
\label{bst2}
  b_{st2} & = &
       - s_a
          /4      \; ;
\end{eqnarray}
\begin{eqnarray}
\nonumber
  b_{ts0} & = &  \{
   ((m_3^2 + s - s_a) m_2^2
   + (s + s_a) m_3^2 - m_3^4) m_1^2
\\
\label{bts0}
& &
  + (m_3^2 - s + s_a) m_2^2 s - m_1^4 m_3^2
     - m_2^4 s - m_3^2 s s_a \}
          /4      \; ;
\\
\label{bts1}
   b_{ts1} & = &  \{
   (m_3^2 - s + s_a) m_1^2
 - (m_3^2 - s - s_a) m_2^2
\\
\nonumber
& &
 + m_3^2 s_a + s s_a - s_a^2\}
          /4      \; ;
\\
\label{bts2}
   b_{ts2} & = &
       - s_a
          /4      \; ,
\end{eqnarray}
                along with coefficients
   $ b_{st2} $, $ b_{st1} $;
   $ b_{ts2} $, $ b_{ts1} $
                one needs to know the dis\-cri\-mi\-nants of
(\ref{D3bst}), (\ref{D3bts}):
\begin{eqnarray}
\label{dstf}
          b_{st}
                  & \equiv  &
                b_{st1}^2 - 4 b_{st2} b_{st0} \\
\nonumber
        & = &
 - (2 (s_a + t_a) m_b^2 - m_b^4 - s_a^2 + 2 s_a t_a - t_a^2)
        \times
\\ \nonumber & &
   (m_1^2 + 2 m_1 m_2 + m_2^2 - s_a)
   (m_1^2 - 2 m_1 m_2 + m_2^2 - s_a)
                / 16 \; ;
\end{eqnarray}
\begin{eqnarray}
\label{dtsf}
          b_{ts}
                  & \equiv  &
                b_{ts1}^2 - 4 b_{ts2} b_{ts0} \\
\nonumber
        & = &
 - (2 (s + s_a) m_3^2 - m_3^4 - s^2 + 2 s s_a - s_a^2)
        \times
\\ \nonumber & &
   (m_1^2 + 2 m_1 m_2 + m_2^2 - s_a)
   (m_1^2 - 2 m_1 m_2 + m_2^2 - s_a)
                / 16 \; .
\end{eqnarray}

\subsection{ Two--variable matrix form of $ D_{3a} $ }

                It is again convenient to determine
                the Gram determinant
        $ D_{3a} $
                defined by eq.
(\ref{D3f})
             as expansion in its variables
        $ s_a, t_a $:
\begin{eqnarray}
\label{D3f2}
             D_{3a}
 & = & \sum_{\alpha_s,\alpha_t}
               A_{\alpha_s,\alpha_t}
                                    s_a^{\alpha_t} t_a^{\alpha_t}
            \; .
\end{eqnarray}
             Explicit expressions for coefficients are
\begin{eqnarray}
\nonumber
A_{0 0} & = & ((((m_a^2 + s) - m_b^2) m_3^2
               - m_3^4 - m_a^2 s) m_b^2)/4 \; ; \\
\nonumber
A_{0 1} & = & ((m_b^2 - m_a^2 + s) m_3^2 + m_b^2 s
            + m_a^2 s - s^2)/4 \; ; \\
\nonumber
A_{1 0} & = & ((m_b^2 + m_a^2 - s) m_3^2
            + m_b^2 m_a^2 - m_a^4 + m_a^2 s)/4 \; ; \\
\nonumber
A_{0 2} & = & ( - s)/4 \; ; \\
\nonumber
A_{1 1} & = & ( - (m_b^2 - m_a^2 - s))/4 \; ; \\
\label{axx}
A_{2 0} & = & ( - m_a^2)/4
 \; .
\end{eqnarray}
                The matrix
        $ \hat{A} $
                of the quadratic form
(\ref{D3f2})
\begin{equation}
\label{D3A}
         D_{3a} =
                  \left(
\begin{array}{c}
                   s_a \\
                   t_a
\end{array}
                             \right)^{\rm T}
                \cdot \hat{A} \cdot
                 \left(
\begin{array}{c}
                   s_a \\
                   t_a
\end{array}
                             \right)
                + A^{\rm T}  \cdot
                 \left(
\begin{array}{c}
                   s_a \\
                   t_a
\end{array}
                             \right)
                + A_{00} \; ,
\end{equation}
               where
\begin{equation}
                 A \equiv
                 \left(
\begin{array}{c}
                   A_{10} \\
                   A_{01}
\end{array}
                             \right)
                  \; ; \;
                 \hat{A} \equiv
                 \left(
\begin{array}{cc}
                   A_{20} & A_{11}/2 \\
                   A_{11}/2 & A_{02}
\end{array}
                             \right)
                  \; ,
\end{equation}
                has the determinant which up to a constant
                coincides with the Gram determinant
\begin{equation}
         D_{2c} \equiv - \Delta_2 ( p_a , p_b ) =
                   (s - (m_a + m_b)^2 )
                   (s - (m_a - m_b)^2 )
                                       / 4 \; ,
\end{equation}
                namely,
\begin{equation}
        A_2 \equiv {\rm Det} \ \hat{A} = - D_{2c} / 16
                     \; .
\end{equation}
                The free term
        $A_c$
                of the corresponding to
(\ref{D3A})
                centered form
\begin{equation}
        A_c = A_{00} - {{1}\over{4}} A^{\rm T} \cdot
              \hat{A}^{-1} \cdot A
\end{equation}
                also is expressed via
        $ D_{2c} $:
\begin{equation}
        A_c =  - m_3^2 D_{2c}
            =  - m_3^2
  ( s - ( m_a + m_b )^2 )
  ( s - ( m_a - m_b )^2 ) / 4
                     \; .
\end{equation}

		Let us give for completeness the coordinates
		of geometrical center of the conic section
		determined by the form
        $ D_{3a} $:
\begin{equation}
\label{sac}
                 \left(
\begin{array}{c}
                   s_a^c \\
                   t_a^c
\end{array}
                             \right)
                  \;  =  \;
                - \frac{1}
                       {2}
                 \hat{A}^{-1} \cdot A
                  \;  =  \;
                 \left(
\begin{array}{c}
                   s + m_3^2  \\

                  m_3^2 + m_a^2
\end{array}
                             \right)
		\; .
\end{equation}


\subsection{ Second--order Gram determinants}

                To control absolute bounds of
        $ s_b $,  $ t_b $
                variables
                one
                should add
                to the list of determinants
        $ D_{2c} $,  $ D_{2a} $,
        $ D_{2as} $,  $ D_{2at} $
\begin{eqnarray}
\label{D2cc}
D_{2c} & \equiv & - \Delta_2 ( p_a, p_b ) \\ \nonumber
       & = & (m_b^2 + 2 m_b m_a + m_a^2 - s)
             (m_b^2 - 2 m_b m_a + m_a^2 - s)/4 \, ; \\
\label{D2aa}
D_{2a} & \equiv & - \Delta_2 ( q_1, q_2 ) \\ \nonumber
       & = & (m_2^2 + 2 m_2 m_1 + m_1^2 - s_a)
             (m_2^2 - 2 m_2 m_1 + m_1^2 - s_a)/4 \, ; \\
\label{D2ss}
D_{2as} & \equiv & - \Delta_2 ( q_1 + q_2, q_3 ) \\ \nonumber
       & = & (m_3^2 + 2 m_3 \sqrt{s} + s - s_a)
             (m_3^2 - 2 m_3 \sqrt{s} + s - s_a)/4 \, ; \\
\label{D2tt}
D_{2at} & \equiv & - \Delta_2 ( p_a, q_3 ) \\ \nonumber
       & = & (m_3^2 + 2 m_3 m_a + m_a^2 - t_a)
        (m_3^2 - 2 m_3 m_a + m_a^2 - t_a)/4 \,
\end{eqnarray}
                already used in previous considerations
                the analogues of
        $ D_{2a} $, $ D_{2as} $,  $ D_{2at} $
                obtained by transposition of particles
        $ 1 \leftrightarrow 3 $,
        $ a \leftrightarrow b $.

                All these determinants
                treated as functions of quantities entering
                their bodies
                are in fact
                represented by the same standard kinematical
                function
        $ \lambda $
                ( cf. eq.
(\ref{lam}))
\begin{equation}
        \Delta_2 ( p, q )
                \; \equiv \;
        {\rm Det}
                 \left(
\begin{array}{cc}
                   p^2 & p \cdot q \\
                  p \cdot q  & q^2
\end{array}
                             \right)
                  \; = \;
               - \frac{1}
                      {4}
               \lambda ( (p + q)^2, p^2, q^2 )
\end{equation}
                with
\begin{eqnarray}
\label{lamsym}
        \lambda ( x , y , z )
                & = &
         x^2 + y^2 + z^2 - 2 x y - 2 y z - 2 z x \\
\nonumber
                & = &
          \left(
                x - ( \sqrt{y} + \sqrt{z} )^2
          \right)
          \left(
                x - ( \sqrt{y} - \sqrt{z} )^2
          \right)                               \\
\nonumber
                & = &
          \left(
                y - ( \sqrt{z} + \sqrt{x} )^2
          \right)
          \left(
                y - ( \sqrt{z} - \sqrt{x} )^2
          \right)                                 \\
\nonumber
                & = &
          \left(
                z - ( \sqrt{x} + \sqrt{y} )^2
          \right)
          \left(
                z - ( \sqrt{x} - \sqrt{y} )^2
          \right)
                     \; .
\end{eqnarray}
                The same applies to
        $ D_{3a} $,
        $ D_{3as} $, $ D_{3at} $
                (and, generally speaking, to
        $ D_4 $
                as well) ---
                there are many forms
                in which
                the so called kinematical
        $ G $--function
                might be represented
                and many elegant properties of it which
                then follow.
                Having in mind quite utilitarian goal we display
                only few of them in the simplest possible
                terms.
                For
                more details reader can look into
\cite{BK}.

\end{document}